\documentclass[12pt]{article}
\def\mpm{{\it mpm}}
\def\Df{{\mathfrak D}}
\def\kk{{\mathcal K}}
\def\nz{{\mathbb Z}}
\def\nn{{\mathbb N}}
\def\ni{{\mathbb I}}
\def\nq{{\mathbb Q}}

\def\nr{{\mathbb R}}
\def\st{{\it espacio-tiempo}}
\def\ST{{\it Espacio-Tiempo}}
\def\MP{$\mathbb{M\hspace{-1.4 mm}P}$} 
\def\MPm{\mathbb{M\hspace{-1.4 mm}P}}
\def\todu{\begin{array}{c}\to\\ 2:1\end{array}}
\usepackage{amssymb}
\usepackage{amsmath}
\usepackage[spanish]{babel}
\usepackage[dvips]{graphicx}
\newtheorem{axiom}{Axioma}[section] 
\newtheorem{definition}{Definici\'on}[section]
\newtheorem{theorem}{Teorema}[section]
\newtheorem{corollary}{Corolario}[section]

\newtheorem{case}{Caso}[theorem]

\newenvironment{proof}[1][Prueba]{\textbf{#1.} }{\ \rule{0.5em}{0.5em}}
\title{\bf Hacia un necesario cambio en los principios matem\'aticos de la filosof\'{\i}a natural.}
\author{Mauricio Ayala-S\'anchez\footnote{E-mail: mauricio@ayala.as}}
\date{}

\begin{document}
\maketitle

\abstract{Siendo las matem\'aticas un lenguaje natural al Hombre y a la f\'{\i}sica, este debe ser constantemente adaptado a nuestras necesidades y nuestra percepci\'on natural. Luego, los conceptos matem\'aticos no son absolutos a la {\it realidad}. Si bien las teor\'{\i}as matem\'aticas son construcciones de la mente, y la {\it existencia} de los objetos en las teor\'{\i}as es un tema de consistencia, o {\sf coherencia} de la teor\'{\i}a (lo que le permite tener plena libertad de estudio), en la f\'{\i}sica las estructuras matem\'aticas no solo deben ser {\sf coherentes}, estas deben tener un referente mucho m\'as tangible a la {\it realidad}. Por lo tanto cualquier modelo matem\'atico no necesariamente es apropiado para la f\'{\i}sica. Entonces qu\'e sentido tiene en la f\'{\i}sica el concepto de punto?, noci\'on elemental para representar los objetos de la {\it realidad}. 

Si tenemos en cuenta que en un buen modelo de la {\it realidad}, la existencia de un objeto matem\'atico con un referente f\'{\i}sico deber\'{\i}a ser equivalente a la posibilidad de su construcci\'on, percepci\'on y comprobaci\'on de existencia f\'{\i}sica, donde la refutaci\'on de no-existencia no significa necesariamente que es posible encontrar una prueba de existencia. Entonces, se obtiene que el lenguaje matem\'atico se vuelve m\'as rico en estructuras, donde el concepto de punto es reemplazado por conceptos m\'as intuitivos y perceptibles. Por un lado, esto posiblemente nos puede acercar a una f\'{\i}sica sin infinitos, y sin paradojas. y por otro lado se restringe el uso de las matem\'aticas y el uso del experimento mental en la f\'{\i}sica.
} 

\newpage
Antes de comenzar, deseo advertir al lector que esta, m\'as que una discusi\'on t\'ecnica, es una discusi\'on desde la conciencia de nuestra percepci\'on\footnote{Esta palabra se usar\'a en un sentido amplio, seg\'un el lector lo considere m\'as apropiado}. Entonces voy a construir este discurso con los {\it s\'{\i}mbolos} m\'as b\'asicos que los recursos me proporcionan, y no empezando con un lenguaje t\'ecnico y elaborado, propio de la f\'{\i}sica.

Teniendo en cuenta que la relaci\'on {\it Matem\'aticas $\&$ F\'{\i}sica}, la definici\'on y representaci\'on del \st\ y de {\it la m\'{\i}nima parte de materia} (\mpm) ha suscitado tantas variantes\footnote{e.g. teor\'{\i}a de cuerdas, geometr\'{\i}a no-conmutativa, conjuntos causales, etc..} en la investigaci\'on f\'{\i}sica (que esto podr\'{\i}a llevar a la desesperaci\'on a cualquier novato que desee abordar estos temas), me pregunto (suponiendo que todas son representaciones posibles) ¿cual es la representaci\'on correcta?, o por lo menos ¿cual es la m\'as adecuada para la f\'{\i}sica?. Si el experimento logra discernir esto, problema solucionado. Pero si no es as\'{\i}, entonces ¿cual es la representaci\'on m\'as general?, dado que al encontrarla, nos podemos concentrar en esta con el indicio de que la f\'{\i}sica est\'a all\'{\i} ``en alg\'un lado''\footnote{Aunque todo parece indicar que es la T. de Cuerdas la soluci\'on m\'as general.}. Sin embargo esta pregunta no ser\'a de nuestro inter\'es, ya que resolverla requerir\'{\i}a de mucho lenguaje t\'ecnico y elaborado, lo cual va en direcci\'on contraria a nuestra premisa inicial. Por lo tanto, la \'unica opci\'on que me queda es 

\begin{quote}
cuestionar (sin llegar a la exageraci\'on) el objeto m\'as b\'asico que se puede percibir de cada esquema, de tal manera que podamos construir una re-presentaci\'on que sea ``la que mejor se aproxime a nuestra concepci\'on de la {\it realidad}''.
\end{quote}

Esto ya presenta una dificultad, porque a pesar de que el lenguaje intenta hacer un concenso,  ``la que mejor se aprox....'' puede tener m\'ultiples interpretaciones. Por ejemplo, podriamos escoger la mejor opci\'on por el consenso o aceptaci\'on que la comunidad cient\'{\i}fica muestre, o la que m\'as se acerque a un lenguaje de simetr\'{\i}as, o por su simplicidad, o dado el caso por su riqueza en estructuras. Y a pesar de que la f\'{\i}sica y en general las ciencias b\'asicas se presentan como un m\'etodo sistem\'atico de abordar y clasificar la {\it realidad}, la investigaci\'on cient\'{\i}fica no deja de ser un arte (no hay un m\'etodo general y sistem\'atico que nos permita abordar y solucionar un problema) y encontrar respuestas es casi {\it serendipity}.

\begin{flushright}
{\it El lenguaje como una acci\'on positiva.} \bf{(I)}
\end{flushright}

Como tal la construcci\'on del lenguaje y su empleo ser\'a siempre una acci\'on positiva, se cree que se puede hablar de algo, lo cual es aun m\'as dificil cuando del lenguaje se quiere hablar\footnote{Aqu\'{\i} no me preocupo del lenguaje y su aprendizaje en el ser humano, sin\'o de su relaci\'on con el individuo}. As\'{\i} como en toda discusi\'on se {\it asume} (o se {\it impone}) una acci\'on primaria o una premisa no demostrable, en todo lenguaje formal se {\it asume} (o se {\it impone}) un primer {\it s\'{\i}mbolo}\footnote{el lector puede darle significado a esta palabra (en cursiva) en un sentido amplio y seg\'un pueda comprender mejor este texto} que no puede ser definido por otros, y cualquier intento de hacerlo puede volverse c\'{\i}clico. En un lenguaje formal los primeros s\'{\i}mbolos con los que nos encontramos es con {\it el objeto} y luego con {\it la relaci\'on} entre objetos (o de otra forma {\it el sujeto} $\&$ {\it el predicado}, o {\it del que se habla} $\&$ {\it lo que se habla}).

Despu\'es de esto, la acci\'on primaria del lenguaje formal es {\sf separar}, {\sf clasificar-conectar} y {\sf representar}, para esto la primera {\it relaci\'on} que suponemos es la identidad, de tal manera que {\it un objeto solo puede ser id\'entico a si mismo}. Esto ya nos permite {\sf separar} o distinguir objetos, si estos no son mutuamente id\'enticos. Pero nuestras acciones son finitas y nuestros recursos limitados, entonces necesitamos descomponer nuestro ``objeto de observaci\'on'' en objetos primarios: debemos {\sf clasificar-conectar}. Para esto distinguimos y seleccionamos lo que es com\'un a las cosas, lo que ``est\'a'' all\'{\i} y all\'a, debemos {\it relacionar}, escojer una particularidad que percibimos de la {\it realidad} (ej. el color, sonido, las cualidades, etc.). Esto nos lleva a la idea del concepto; primer indicio evidente de la abstracci\'on en el lenguaje, lo cual no es una exclusividad de las ciencias ``duras''. 

La necesidad primaria del lenguaje es {\it comunicar}, expresar algo, lo cual no es posible si no tenemos un concenso con nuestro interlocutor, para esto acompa\~nado a la idea de concepto, est\'a la acci\'on de {\sf representar}. En nuestro caso la primera representaci\'on conceptual es auditiva: un sonido, un gemido, el cual est\'a en corres-pondencia con nuestra percepci\'on. En un lenguaje m\'as elaborado, la representaci\'on tambi\'en tiene un referente visual: un s\'{\i}mbolo, una forma visual. Luego cualquier construcci\'on o creaci\'on humana no puede ser en sus {\it objetos} primarios ajena a nuestros sentidos. Valoramos, clasificamos, extraemos, correjimos siempre en conexi\'on directa con nuestra percepci\'on: no es posible concebir un tri\'angulo rect\'angulo si nuestros sentidos no pueden indic\'arnoslo (en particular la visi\'on). Solo despu\'es en jerarqu\'{\i}a de abstracci\'on, podemos usar conceptos que no tienen un referente directo con los objetos de la {\it realidad}. 

Por consiguiente, en resumen, tenemos que la acci\'on efectiva del lenguaje es conectar o relacionar conceptos, esta es una construcci\'on m\'ultiple entre nuestra percepci\'on, nuestro pensamiento y la expresi\'on. 

\begin{flushright}
{\it Las matem\'aticas como un lenguaje natural al Hombre.} \bf{(II)}
\end{flushright}

El lenguaje es una competencia del Hombre, el cual es un ser racional en la medida que {\sf separa}, {\sf clasifica-conecta} y {\sf representa}, es decir en la medida que este abstrae \footnote{por ejemplo un pintor no deja de serlo porque usa ``el pincel'' y ``la pintura'', igual este usa un ``alfabeto'': el color, as\'{\i} como el m\'usico usa la escala, el escultor: la forma, etc.}. Y solo se realiza en un ser matem\'atico en el momento que valora, ordena o jerarquiza. 

La primera valoraci\'on con la que nos topamos es con la de dos estados: $\{\top,\perp\}$, $\{si,\ no\}$, $\{\uparrow,\downarrow\}$, etc. que no pueden ser lo mismo. Por ejemplo algo no puede ser falso ($\perp$) y verdadero ($\top$) al mismo tiempo, esta es una exijencia de {\sf coherencia}, y es el primer criterio que nos permite asimilar y reconocer nuestro entorno. Pero no el \'unico, ya que ser\'{\i}a absurdo clasificar la {\it realidad} en dos estados, de la misma forma que {\it blanco $\&$ negro} son insuficientes para describir lo que nuestros ojos observan. La primera costrucci\'on matem\'atica es una acci\'on l\'ogica (lo cual es incidental y no esencial a las matem\'aticas). Luego, la {\it existencia} de un objeto en una estructura matem\'atica es un tema de consistencia, o {\sf coherencia} en la teor\'{\i}a.

De manera similar a cualquier lenguaje, las matem\'aticas se costruyen imponiendo un alfabeto, y un conjunto de premisas no demostrables (axiomas), de tal manera que cualquier construcci\'on o afirmaci\'on posterior (teoremas, lemas, corolarios, etc.) se deduce de estas premisas iniciales, siguiendo un n\'umero de pasos finitos seg\'un las reglas l\'ogicas impuestas. A diferencia de cualquier otro lenguaje, en las matem\'aticas se pueden estudiar estructuras que no tienen necesariamente un referente real o perceptible (aunque inicialmente lo tuvieron), y que pueden solo ser juegos de la mente humana. Es en esta medida que las matem\'aticas son en conjunto un lenguaje natural al Hombre y es la forma como reconocemos el {\it mundo}, sin embargo decir que {\it la natura} es matem\'atica, es decir que cada parte del universo se relaciona con su entorno usando un lenguaje, lo cual es ya muy antropom\'orfico y egoc\'entrico, esto no tiene que ser as\'{\i}. Cualquier comentario no deja de ser una valoraci\'on que solo le compete al Hombre: Es el Hombre quien pregunta y es el Hombre quien responde. Luego, no hay conocimiento fuera del Hombre (dado que el conocimiento est\'a inmerso en el lenguaje, el cual solo le compete al Hombre).

\begin{flushright}
{\it El n\'umero como un cuantificador.} \bf{(III)}
\end{flushright}

Despu\'es de $\{0,1\}$ el individuo se enfrenta con el n\'umero, por la necesidad de clasificar una propiedad de cantidad, de ocupaci\'on. Construimos los Naturales ($\nn$) por agrupaci\'on (+) de objetos, y por sustracci\'on (-) construimos los Enteros ($\nz$): puedo decir que {\it tengo} 1000 pesos en el banco, sin embargo qu\'e sentido tiene decir que {\it tengo} -500 pesos en el banco, si realmente no tengo nada. Entonces el s\'{\i}mbolo - surge aqu\'{\i} por la necesidad de decir que le debo al banco 500 pesos. Introduzco los Racionales ($\nq$) porque necesito partir, dividir. Luego los Reales ($\nr$) por {\sf coherencia} geom\'etrica, Los Imaginarios porque necesitamos n\'umeros que su cuadrado sea nega-tivo ($x^2<0$), despues los cuaterniones, octoniones, sedeniones, etc. para satisfacer alguna propiedad algebr\'aica en particular. Aqu\'{\i} vemos que la idea del n\'umero siempre se ha visto adaptada por la necesidad de decir algo o de completar una propiedad. Sin embargo, es $\nr$ el objeto que ocupa un papel central en las Matem\'aticas standard, ya que las otras o son subconjuntos o son extensiones basadas en $\nr$, esto teniendo en cuenta el {\sf Teorema fundamental sobre n\'umeros finitos}\footnote{Teorema: {\it Cada n\'umero finito es aproximado por un \'unico n\'umero real}. {\it Prueba:} Dado $t$ un n\'umero finito cualquiera y $H=\{y|y\in \nr,\ y<t\}$. Por el teorema de completes para $\nr$, $H$ tiene una m\'{\i}nima cota superior, etiquetada por $a$, por consiguiente $a$ aproxima a $t$. Si esto no es as\'{\i}, entonces existe un n\'umero positivo real $h$ tal que $h<|t-a|$. Por lo tanto, hay justo dos casos, si $a\neq t$
\begin{enumerate}
\item Asuma que $a<t$, luego $h<t-a$, o $h+a<t \Rightarrow h+a\in H$, pero $a<a+h$, entonces $a$ no es una cota superior para $H$.
\item Asuma que $a>t$, luego $h<a-t$, o $t<a-h \Rightarrow a-h$ es una cota superior para $H$, pero $a>a-h$, entonces $a$ no es la menor cota superior de $H$.
\end{enumerate}
Esto prueba que $a\simeq t$. Por supuesto, si un n\'umero real es aproximado por dos n\'umeros reales (ej. $0.9\bar{9}=1$), entonces la diferencia entre estos n\'umero es un n\'umero infinitesimal, que en los reales standard este n\'umero solo puede ser cero, dado que la l\'ogica impuesta all\'{\i} es una l\'ogica bivaluada, es decir, para cualquier n\'umero real $x$, se tiene que $x=0$ o $x\neq 0$.} 

Despu\'es de esto han surgido nuevos n\'umeros, por ejemplo las variables de Grassmann ($x^2=0$) porque necesitamos una f\'{\i}sica sin infinitos, luego los infinitesimales de Robinson para comprender el continuo de la recta de la misma forma como percibimos el continuo fluir del tiempo. Los \'ultimos n\'umeros en aparecer en la arena, son los n\'umeros nilpotentes de la geometr\'{\i}a sint\'etica ($x^n=0$ para alg\'un $n>0\in \nn$) como otra alternativa al continuo. 

Tanto los naturales, como las dem\'as estructuras de n\'umeros han surgido para satisfacer una construcci\'on mental. Por ejemplo, en la f\'{\i}sica nadie coloca en duda el uso de los n\'umeros complejos, en particular en la Mec\'anica Cu\'antica (MC), dado que satisfacen un criterio algebr\'aico, adem\'as son c\'omodos para escribir las ecuaciones. Sin embargo es realmente muy dificil indagar que realidad tienen los n\'umeros fuera de nuestra mente. Construimos el lenguaje, en particular el lenguaje matem\'atico por la necesidad de reconocer nuestro entorno, no podemos decir m\'as y cualquier otra respuesta es necedad.

\begin{flushright}
{\it La f\'{\i}sica y sus conectivos con las matem\'aticas.} \bf{(IV)}
\end{flushright}

Las ciencias f\'{\i}sicas pueden ser definidas como ``una sistematizaci\'on del conocimiento obtenido por la medici\'on'', y es una convenci\'on que este conocimiento sea formulado como una descripci\'on de nuestra {\it realidad} llamada ``Universo F\'{\i}sico''. Similar al conocimiento matem\'atico, en la f\'{\i}sica se parte de premisas iniciales, principios b\'asicos o {\it leyes f\'{\i}sicas} (por ej. la causalidad, la suavidad del \st, las leyes de conservaci\'on, la invariancia de la velocidad de la luz, etc.) que nos permitan deducir consecuencias que son contrastables. Sin embargo estas {\it leyes f\'{\i}sicas} no son inherentes al ``mundo'', son impuestos por nosotros para reconocer los diferentes elementos de la {\it realidad}. Y aunque no tienen un referente tangible y/o evidente, son aceptados porque sus consecuencias no entran en aparente choque con el experimento y proporcionan un descripci\'on conceptual adecuada a la {\it natura}. Pero dado el momento que alg\'un principio o postulado entre en contradicci\'on con nuestra {\it realidad}, nadie dudar\'a en removerlo de los postulados f\'{\i}sicos. Y el hecho de que sus consecuencias sean corroborables no significa que estos principios sean ``ciertos''\footnote{Recuerde que, de algo falso se puede obtener una certeza, por ej. $((2=3)\Rightarrow (3=2))\Rightarrow (2+3=3+2=5). $}, estos se aceptan por una conveniencia conceptual. 

Luego, de la misma forma que en matem\'aticas se imponen unas reglas l\'ogicas de deducci\'on y una sistematizaci\'on del conocimiento, en la f\'{\i}sica tambi\'en se tiene esta estructura. Por lo tanto, la f\'{\i}sica es un conocimiento matem\'atico de la {\it natura}, o de otra forma las matem\'aticas es el lenguaje natural de la f\'{\i}sica. Pero a diferencia de las matem\'aticas en la f\'{\i}sica s\'{\i} hay una necesidad de tener objetos matem\'aticos con un referente tangible o por lo menos que sus consecuencias sean perceptibles a nuestra {\it realidad}. 

Siendo esta {\it realidad} compleja, para entender las relaciones mutuas entre sus ele-mentos, debemos separar la experiencia en objetos simples de estudio, o conceptos b\'asicos que: 
\begin{enumerate}
\item pueden ser no medibles y no perceptibles (definibles) (conceptos anal\'{\i}ticos: diferenciabilidad, el punto, el espacio infinito, etc.), 
\item pueden ser medibles y no perceptibles (definibles) (propiedades o relaciones: la masa, la carga, etc.), 
\item pueden ser medibles y perceptibles (definibles) (objetos de la experiencia: la luz, etc.).
\end{enumerate}

Nuestra concepci\'on de la materia est\'a asociada a la interpretaci\'on conceptual de la medici\'on, primero hacemos una construcci\'on mental sobre lo que pensamos que vamos a medir, dise\~namos un proceso que nos lleve a su medici\'on, y luego expresamos en s\'{\i}mbolos matem\'aticos, lo que pensamos que estamos haciendo cuando medimos cosas. Por esto, si no tenemos una concepci\'on de lo que estamos midiendo, el resultado no puede persuadirnos a creer algo en particular. Entonces, todos nuestros resultados son derivados de la interpretaci\'on conceptual sobre la cual colocamos los resultados de la medici\'on, y que debe ser consistente con nuestra interpretaci\'on del proceso de medici\'on. Por lo tanto definimos s\'{\i}mbolos con propiedades que est\'an en correspondencia con la concepci\'on introducida. 

\begin{flushright}
{\it El n\'umero en la medici\'on.} \bf{(V)}
\end{flushright}

Los datos de la f\'{\i}sica son mediciones, pero no podemos construir algo de solo mediciones sin un preconcepto del objeto de medici\'on ({\it medible}) y las circunstancias a las que se refiere, para esto debemos encontrar una conexi\'on con la medici\'on:
\begin{center}
{\it Todo observable f\'{\i}sico es un observable espacial, con un referente temporal.}
\end{center}
De otra manera, toda medici\'on f\'{\i}sica, sin importar que se est\'a midiendo, esta siempre se hace con referencia a una escala en el \st, ya sea para medir la cantidad de masa en una balanza o la carga de un electr\'on. Por lo tanto cualquier propiedad que se estudie, es una propiedad f\'{\i}sica si esta se puede representar en el \st, de otra forma cualquier propiedad tiene contenido f\'{\i}sico si afecta la din\'amica de un sistema f\'{\i}sico. 

Ahora podemos inferir las reglas b\'asicas de la medici\'on:
\begin{enumerate}
\item Un observable es la extensi\'on entre {\it dos entradas f\'{\i}sicas} o dos ``marcaciones''.
\item La medida es la proporci\'on entre dos observables (medida standard $\&$ medida casual).
\item Los observables son aditivos. 
\item Un {\it medible} tiene una medida, sin embargo cualquier n\'umero de {\it medibles} pueden tener la misma medida. 
\item Un {\it medible} determina completamente la medida, pero no al contrario. 
\end{enumerate}
Visto con m\'as detalle, siempre que se hace una medida, nosotros solo observamos posiciones y velocidades relativas, luego un observable de coordenada o momentun siempre involucra dos entradas f\'{\i}sicas. Pero una medici\'on siempre involucra cuatro entradas f\'{\i}sicas: dos correspondientes al observable que dice que se va a medir y dos al observable de comparaci\'on usado como standard o medida patr\'on $\ell$, el cual se construye tomando la extensi\'on entre dos entradas f\'{\i}sicas o marcaciones establecidas. Por ejemplo en una medici\'on de distancia, la extensi\'on entre dos objetos es comparada con la extensi\'on entre dos graduaciones marcadas sobre una escala. Por consiguiente, cualquier medida casual es un m\'ultiplo racional de $\ell$ (m\'as exactamente un m\'ultiplo natural de $\ell$), de otra manera cualquier proceso de medici\'on necesita de cuatro entradas o ``marcaciones''. 

Acompa\~nado al proceso de medici\'on hay un error ($\pm \ell/2$) que nos indica la incertidumbre o indistinguibilidad en la medida, de tal manera que la precisi\'on se puede aumentar dividiendo $\ell$ en $n$ partes iguales. Esto nos permite una nueva escala de medici\'on que se puede mejorar seg\'un el desarrollo tecnol\'ogico nos lo permita. Luego, cualquier elemento del conjunto de las mediciones $\Delta$ se reporta como $x\ell$, de tal mane-ra que el conjunto $K$ de valores $x$ tiene la propiedad de ser un campo algebraico\footnote{$K=\{x|x\ell\in \Delta\}$ es un campo algebraico, si $K$ tiene definida las operaciones de {\it suma} (+) y {\it multiplicaci\'on} (.) con la propiedad de grupo abeliano cada una, adem\'as de una ley distributiva: 
$$
(x+y)\ell=(y+x)\ell,\quad (x+(y+z))\ell=((x+y)+z)\ell,\quad (x+0)\ell=x\ell,\quad (x+x')\ell=0\ell
$$
$$
(x.y)\ell=(y.x)\ell,\quad x.(y.z)\ell=(x.y).z\ell, \quad x.1\ell=x\ell,\quad (\mbox{si }x\neq 0,\ \exists y|\ (x.y)\ell=1\ell) 
$$
$$
z.(x+y)\ell=(z.x+z.y)\ell
$$\label{campo-alg}}.

Sin embargo, no es posible deshacernos de la incertidumbre en la medici\'on, y una medida exacta solo se puede alcanzar en el l\'{\i}mite $\ell \to 0$, y por lo tanto en un tiempo infinito, si esto tiene alg\'un sentido f\'{\i}sico. En cierta forma el valor que se obtiene de la medici\'on es solo una cota a nuestra ignorancia (la cual se puede representar proporcional a $\ell$). Aqu\'{\i} ya tenemos dificultades, porque no tiene mucho sentido hablar de un multiplo irracional de $\ell$, luego no es cierto que $\Delta = \nr*\ell=\{r*\ell\ |\ r\in \nr\}$ (o $\Delta/\ell\neq \nr$). 

\begin{flushright}
{\it $\pi$ y la suavidad del espacio-tiempo.} \bf{(VI)}
\end{flushright}

Aunque podemos calcular casi cualquier cosa sin la necesidad de los n\'umeros Irracionales, no podriamos entender ni construir el c\'alculo diferencial o la geometr\'{\i}a sin estos. Como calcular el \'area de un c\'{\i}rculo o darle valor a la hipotenusa de un tri\'angulo equil\'atero?. 

Sin embargo, qu\'e sentido tiene en la f\'{\i}sica los n\'umeros irracionales, en particular qu\'e significa el n\'umero $\pi$ en muchas ecuaciones y par\'ametros f\'{\i}sicos, por ejemplo en la constante de Plank ($\hbar=h/2\pi$) y las relaciones de incertidumbre de Heisenberg, fundamentales en la mec\'anica cu\'antica. La primera referencia que tenemos del n\'umero $\pi$ en las matem\'aticas es el cociente entre el per\'{\i}metro ($P$) de la circunferencia y su di\'ametro ($D=2r$), es decir $\pi=\frac{P}{2r}$. Por otro lado $\pi$ aparece en las ecuaciones cuando hacemos una integraci\'on de \'area o de volumen, o cuando calculamos integrales de flujo. 

\begin{figure}[ht]
\begin{center}
\begin{tabular}{ccccc}
\includegraphics[height=3.0cm,width=3.0cm]{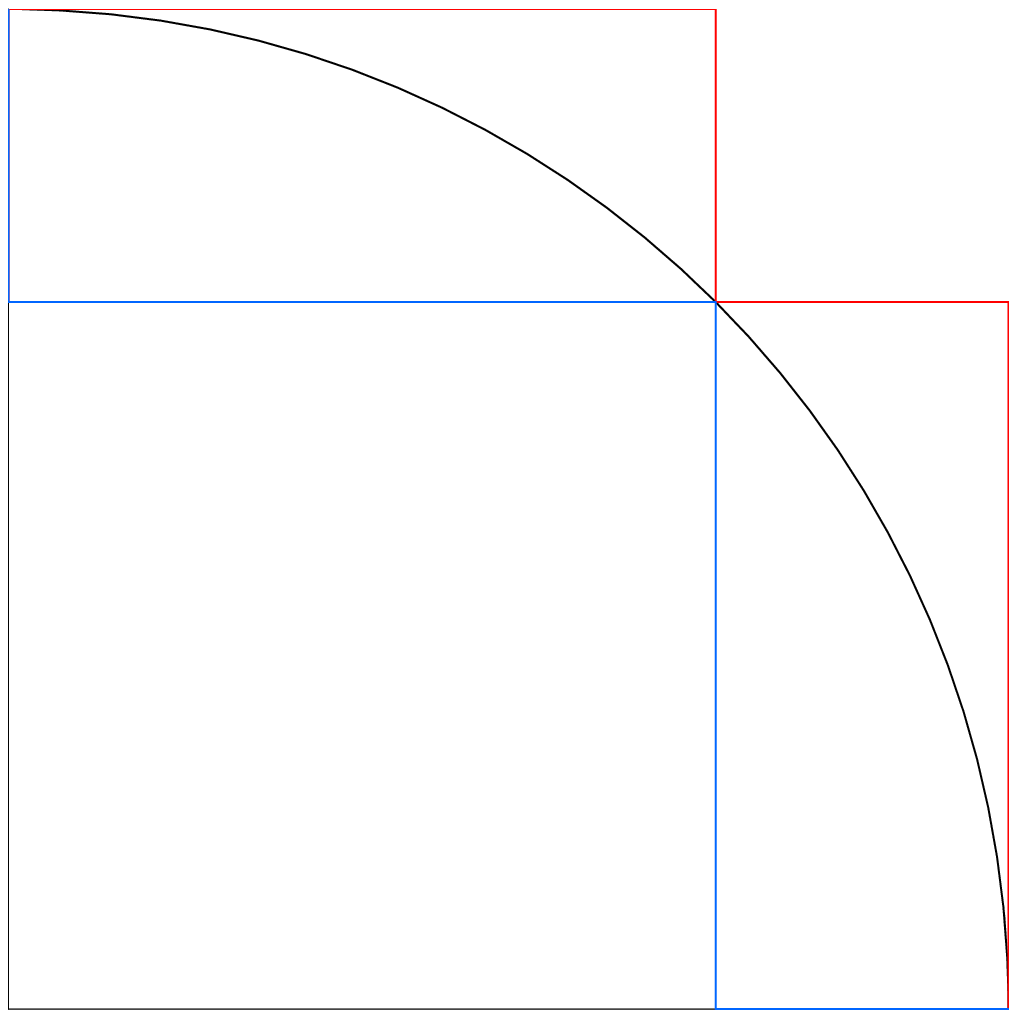}& 
\includegraphics[height=3.0cm,width=3.0cm]{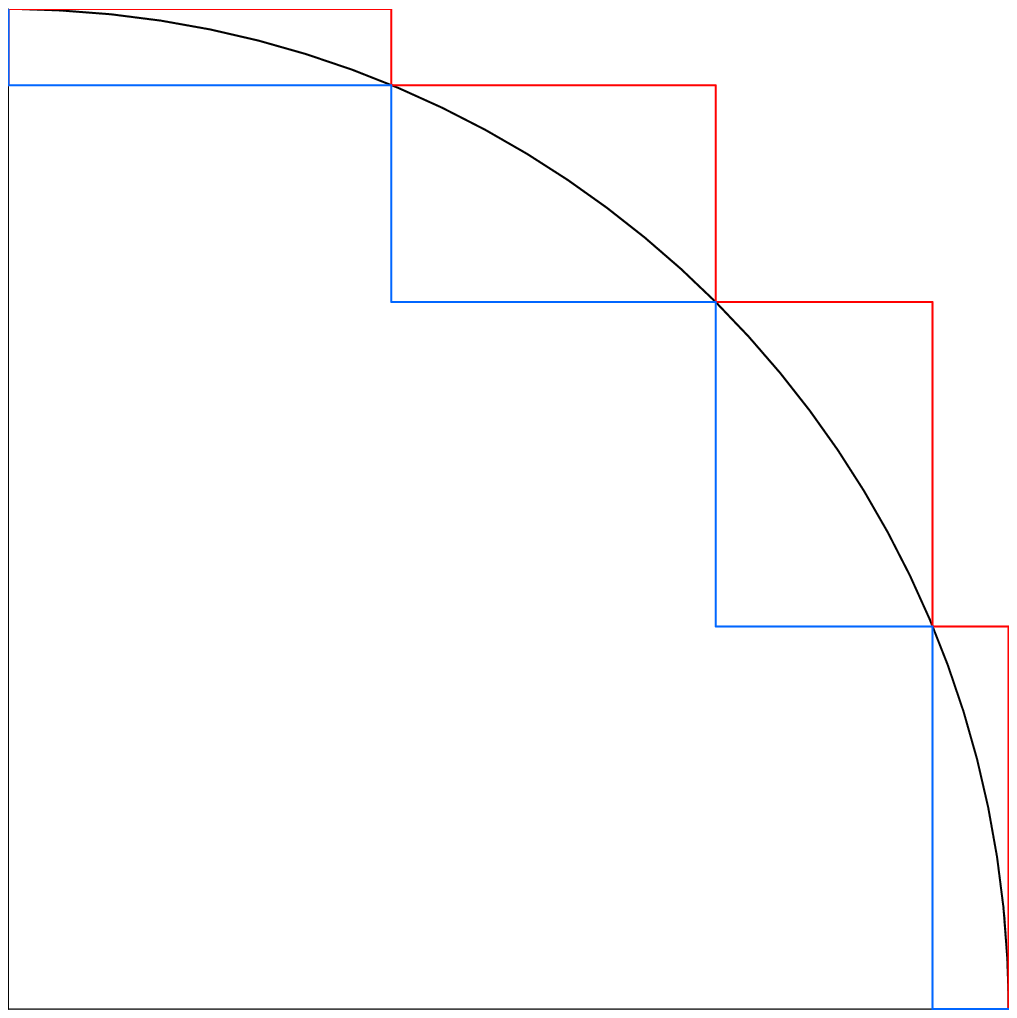}&
\includegraphics[height=3.0cm,width=3.0cm]{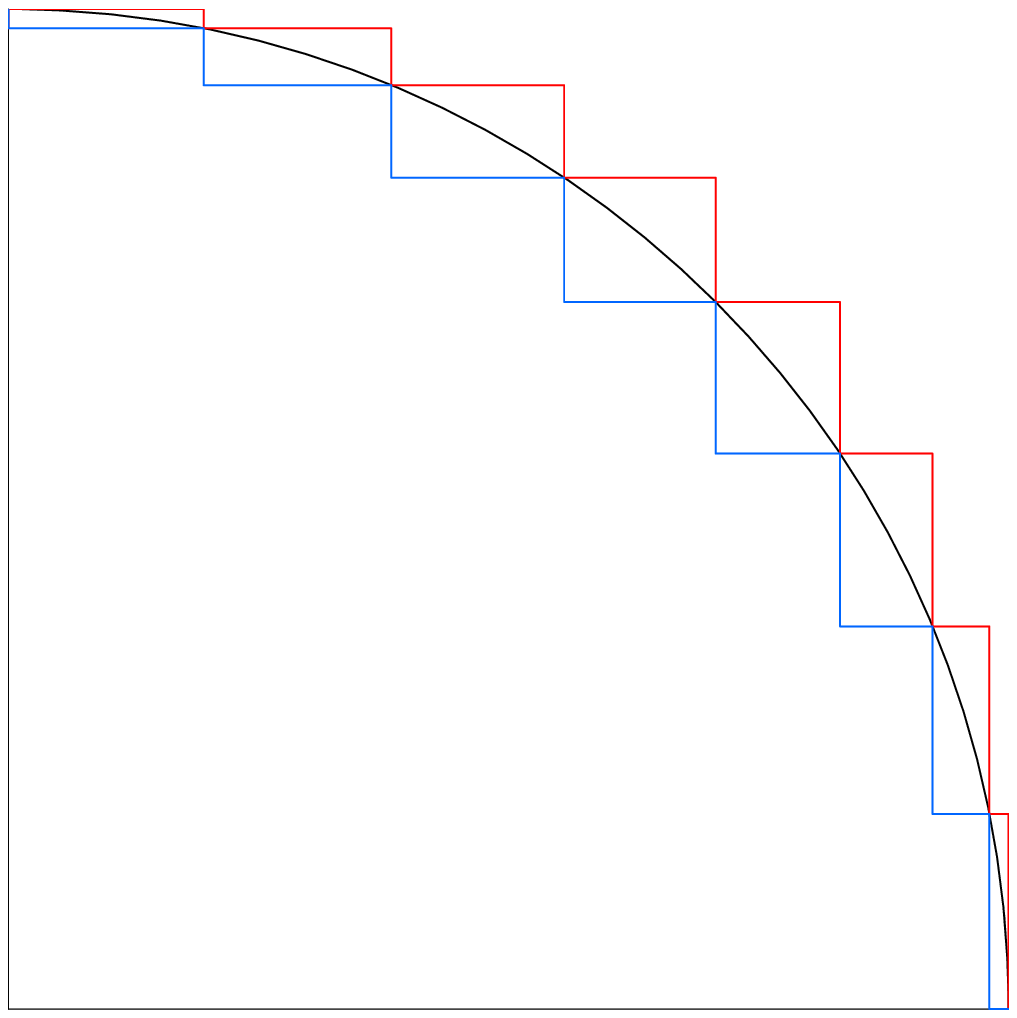}&...& 
\includegraphics[height=3.0cm,width=3.0cm]{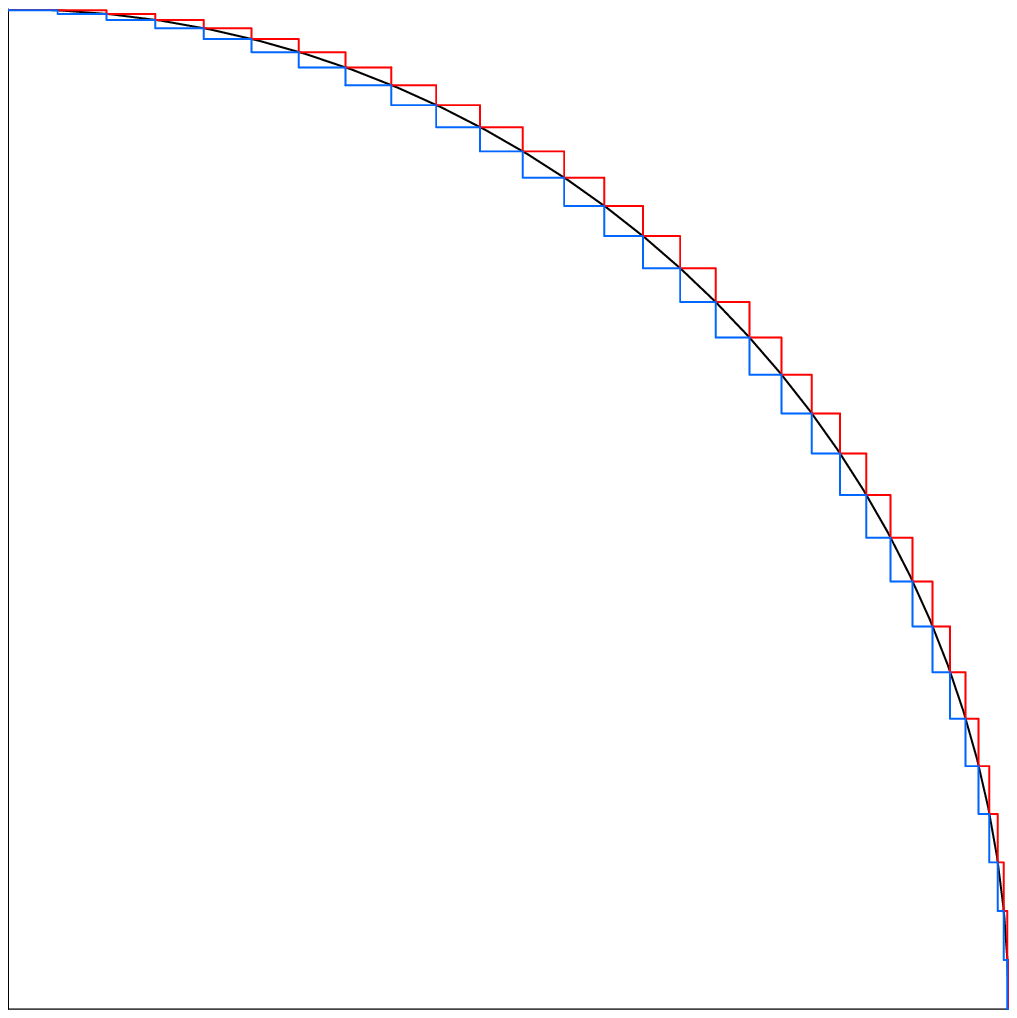}
\end{tabular}
\end{center}
\caption{Aproximaciones sucesivas inscritas $A_n$ (Azul) y circunscritas $R_n$ (Rojo) para la circunferencia.}
\label{graf2}
\end{figure}

Hasta aqu\'{\i} $\pi$ es solo un s\'{\i}mbolo, pero si queremos calcular expl\'{\i}citamente este n\'umero, esto se puede hacer por aproximaciones sucesivas de pol\'{\i}gonos inscritos y circunscritos a una circunferencia, como lo hizo Eudoxo y Arqu\'{\i}medes. Pero nosotros usaremos otro m\'etodo, no usando pol\'{\i}gonos, sin\'o como se indica en la figura \ref{graf2}, similar al esquema geom\'etrico para calcular el \'area bajo una curva cualquiera (por sumas de Riemann). Es decir, siguiendo el contorno de los rect\'angulos inscritos ($A_n$) y circunscritos ($R_n$) en la aproximaci\'on de \'area.

No es dificil calcular la longitud ($\ell$) para el segmento de circunferencia en cada aproximaci\'on.
$$
\ell(A_1)=\ell(R_1)=2r,\quad \ell(A_2)=\ell(R_2)=2r,\quad....\ ,\quad\ell(A_n)=\ell(R_n)=2r,
$$
luego
$$
\lim_{n\to\infty}\ell(A_n)=\lim_{n\to\infty}\ell(R_n)=2r.
$$
Entonces el per\'{\i}metro de la circunferencia toma valor 
$$
P=4*2r=8r\qquad \Rightarrow \qquad \frac{P}{2r}=4\neq \pi\simeq 3.14159...,
$$

\begin{figure}[ht]
\begin{center}
\begin{tabular}{ccccc}
\includegraphics[height=1.5cm,width=3.5cm]{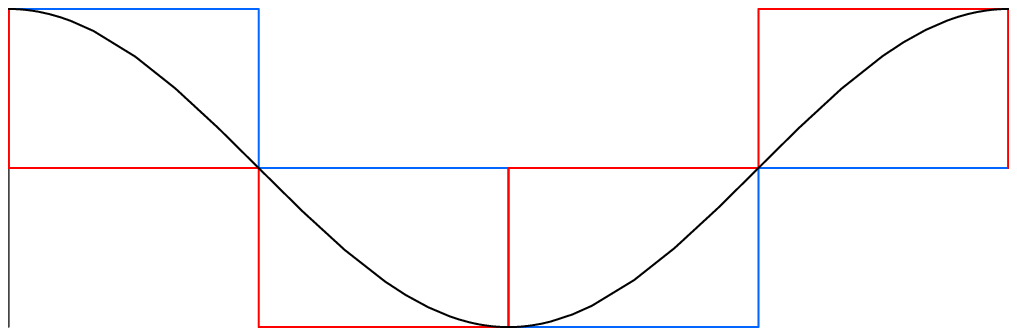}& .....& 
\includegraphics[height=1.5cm,width=3.5cm]{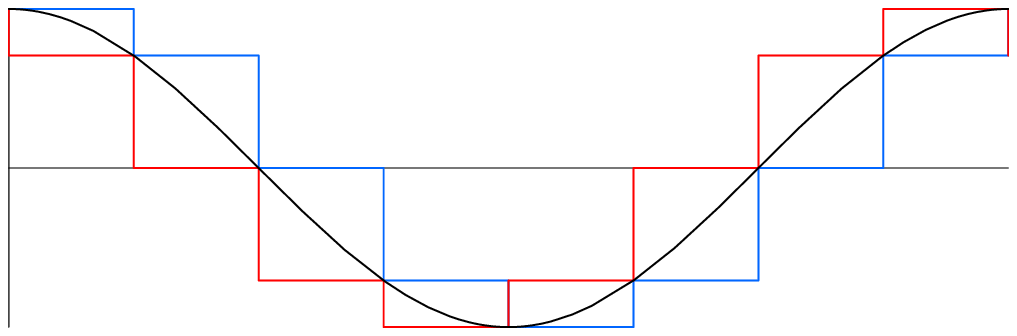}& .....& 
\includegraphics[height=1.5cm,width=3.5cm]{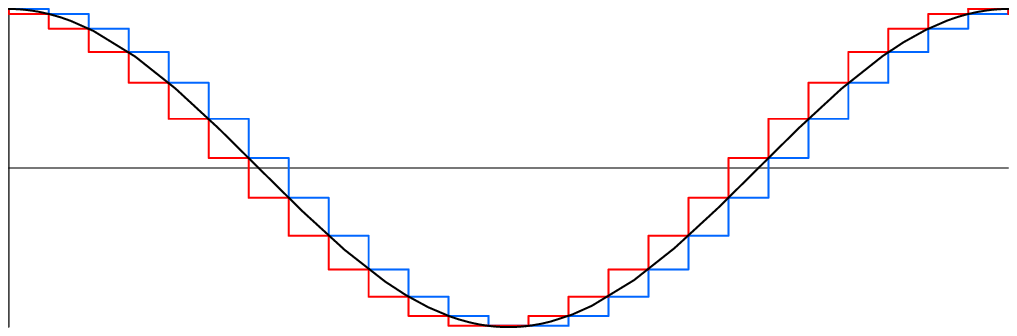}
\end{tabular}
\end{center}
\caption{Aproximaciones sucesivas inscritas $A_n$ (Azul) y circunscritas $R_n$ (Rojo) para $\gamma$.}
\label{graf3}
\end{figure}

Que ha fallado aqu\'{\i}?, por un lado las aproximaciones sucesivas convergen a la circunferencia ($\cal C$), lo que no es dificil de demostrar ya que la convergencia es uniforme\footnote{Dado un $\delta$ puedo encontrar un $m$ tal que para todo $n>m$, $d(A_n,{\cal C})=d(R_n,{\cal C})<\delta$}. Por otro lado el \'area bajo la curva en el l\'{\i}mite $n\to \infty$ toma el mismo valor conocido por el c\'alculo diferencial standard ${\cal A}=\pi r^2$. El problema radica en que el l\'{\i}mite $n\to \infty$, la curva que obtenemos por aproximaci\'on de $A_n$ y $R_n$ no es diferenciable en punto alguno. Esto tambi\'en lo podemos hacer con cualquier curva y obtener una medida diferente a la que se obtiene con el c\'alculo diferencial, por ejemplo, tomese la siguiente curva cosenoidal (Ver la figura \ref{graf3}):
$$
\gamma(t)=(t,\cos(t))\qquad t\in [-\pi,\pi].
$$

\begin{figure}[ht]
\begin{center}
\begin{tabular}{ccccc}
\includegraphics[height=3cm,width=3cm]{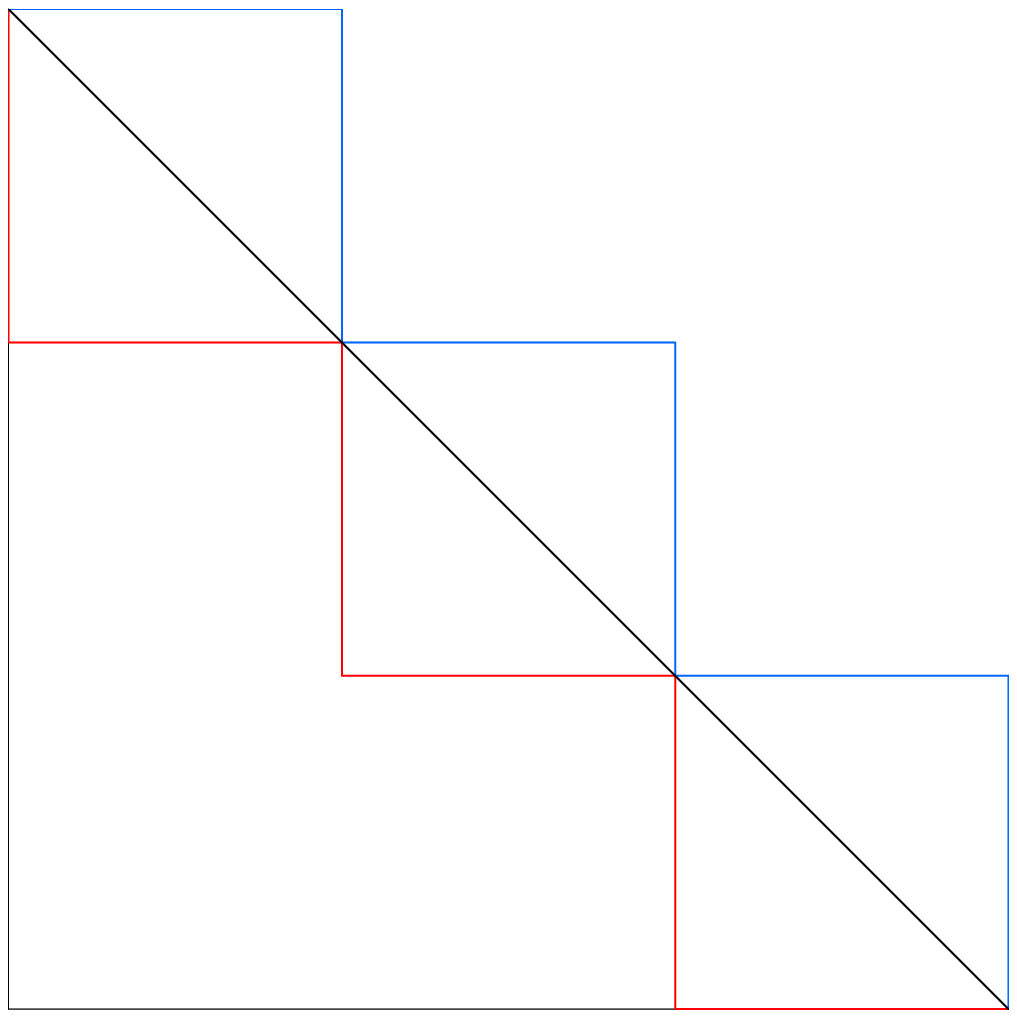}& ...& 
\includegraphics[height=3cm,width=3cm]{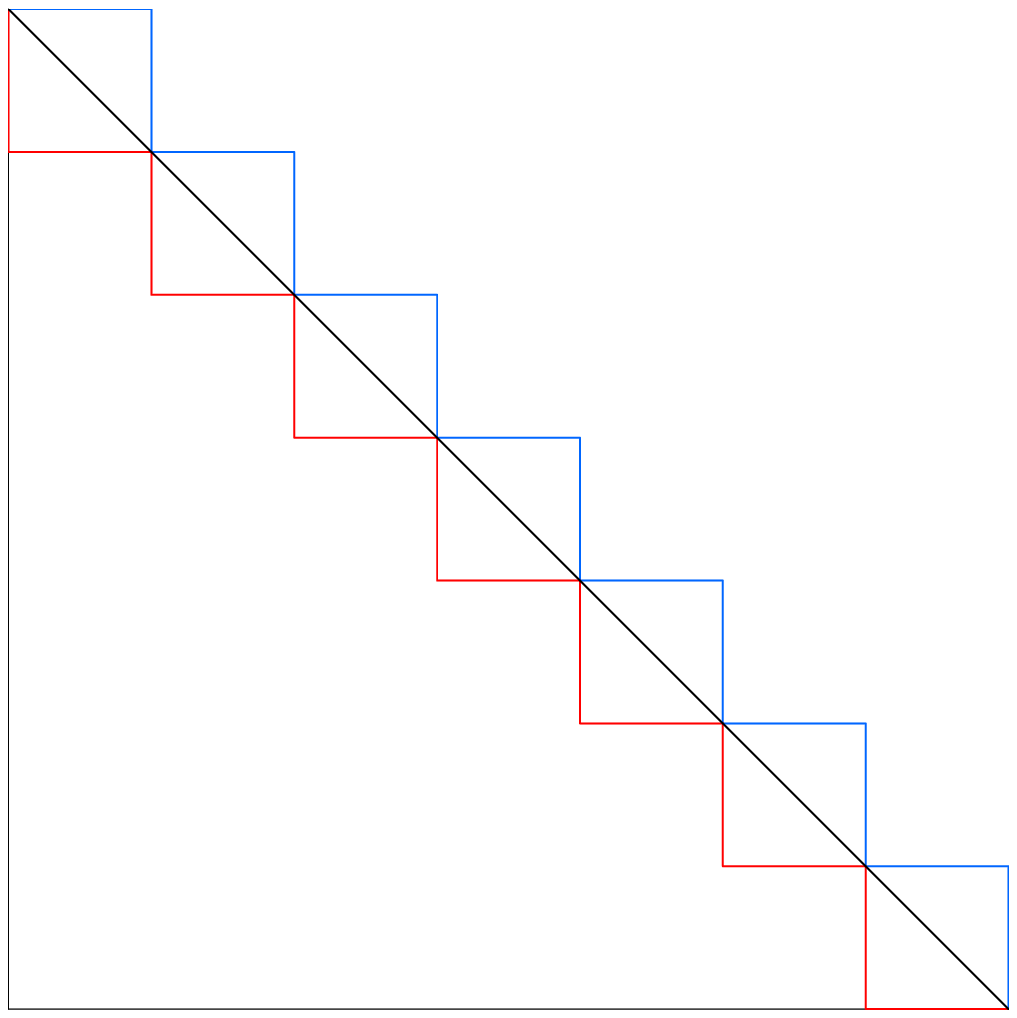}& ...&
\includegraphics[height=3cm,width=3cm]{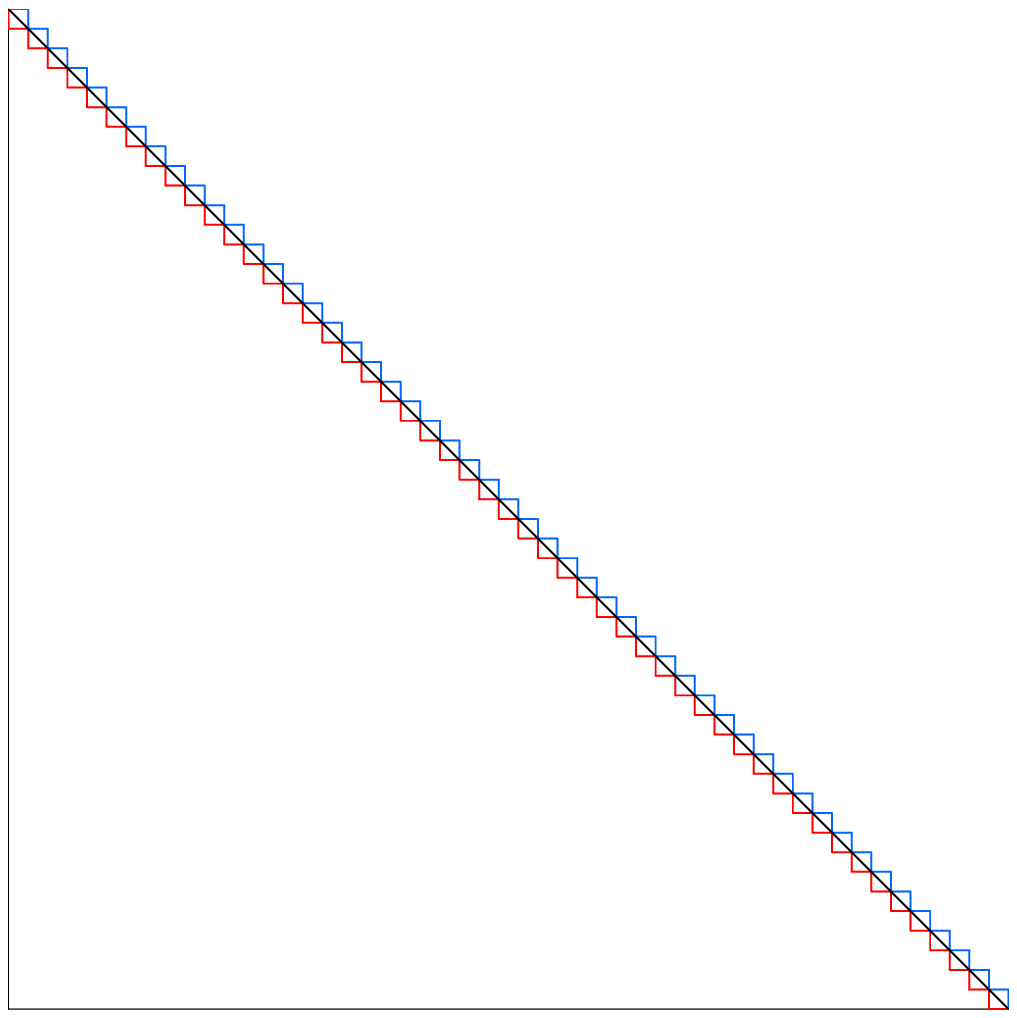}
\end{tabular}
\end{center}
\caption{Aproximaciones sucesivas inscritas $A_n$ (Azul) y circunscritas $R_n$ (Rojo) para la recta.}
\label{graf1}
\end{figure}

La medida que se obtiene para la longitud de la curva es 
$$
\ell(A_1)=\ell(R_1)=\ell(A_2)=\ell(R_2)=....=\ell(A_n)=\ell(R_n)=2\pi+2,
$$
luego
$$
\lim_{n\to\infty}\ell(A_n)=\lim_{n\to\infty}\ell(R_n)=2\pi+2 \simeq 8.2831...,
$$
lo que es diferente de 
$$
\int_0^{2\pi} \sqrt{1 + (\gamma' (t))^2} dt=\int_0^{2\pi} \sqrt{1 + Sin[t]^2} dt \quad \simeq \quad 7.6404.
$$
No es dificil inventarse otras aproximaciones que pueden converger a otros valores de longitud en las anteriores curvas y para cualquier curva suave, y que tienen la particularidad de no ser diferenciables en punto alguno, por ej. la aproximaci\'on diente de sierra, fractales, etc. Como un \'ultimo caso a mostrar, en la figura \ref{graf1} tenemos una aproximaci\'on a la recta, en la cual falla el teorema de Pit\'agoras, dado que 
$$
\lim_{n\to\infty}\ell(A_n)=\lim_{n\to\infty}\ell(R_n)=2L \neq \sqrt{L^2+L^2}=\sqrt{2}L
$$

Como un ejercicio el lector puede comprobar que esto tambi\'en sucede en el {\sf Teorema de los residuos}\footnote{Teorema: \it Si $C$ es un contorno cerrado simple positivamente orientado, dentro del cual y sobre el cual una funci\'on $f$ es anal\'{\i}tica a excepci\'on de un n\'umero finito de puntos singulares $z_k$ ($k=1, 2, ..., n$) interior a $C$, entonces
$$\int_C f(z)dz= 2\pi\sum_{k=1}^n Res_{z=z_k} f(z)$$} en Variable Compleja, donde las integrales pueden converger a cualquier valor si el contorno no es diferenciable en punto alguno, t\'omese por ejemplo la estrella de Kock centrada en el origen sobre la integral de $f(z)=1/z$. 

Por otro lado, el caracter diferencial de las ecuaciones f\'{\i}sicas implican que los fen\'omenos f\'{\i}sicos son governados por leyes que operan con una precisi\'on m\'as all\'a del l\'{\i}mite de verificaci\'on experimental, y aunque todos los resultados obtenidos experimentalmente son descritos con valores finitos en proporci\'on a una escala, la constante aparici\'on del n\'umero $\pi$ (y otros irracionales) en la f\'{\i}sica es un indicio claro de la necesidad de tener un \st\ suave o diferenciable, indispensable para poder hacer c\'alculos y obtener resultados \'unicos, lo cual es fundamental en un modelo f\'{\i}sico.

Si tomamos como valido el {\sf Principio antr\'opico}\footnote{Principio antr\'opico: percibimos el universo tal como es, porque nosotros existimos} y creemos que el universo evolucion\'o desde una condici\'on diferente a la que tenemos actualmente, se obtiene que es necesario considerar el continuo de la recta real para describir el \'algebra de eventos asociados a la evoluci\'on del universo, debido a que esta \'algebra implica un conjunto no-enumerable de eventos. Y aunque este conjunto no es completamente medible, se considera la posibilidad de su existencia como una condici\'on necesaria para comprender el {\sf Principio antr\'opico}. Luego, la causalidad del principio antr\'opico est\'a relacionada directamente con la continuidad del \st\ y la no-enumerabilidad del \'algebra de eventos asociados. Por consiguiente, tenemos otra justificaci\'on de la necesidad de los n\'umeros irracionales en la f\'{\i}sica y por lo tanto de los n\'umeros reales como objeto elemental del lenguaje matem\'atico de la f\'{\i}sica.

\begin{flushright}
{\it El problema longitud $\&$ escala, o diferenciabilidad $\&$ punto.} \bf{(VII)}
\end{flushright}

Aunque la derivabilidad de una estructura matem\'atica standard est\'a definida como el l\'{\i}mite a un punto, el caracter diferencial y puntual del \st\ son dos caracter\'{\i}sticas que deben ser consideradas independientemente. Ya que para definir la variaci\'on de una cantidad f\'{\i}sica no es indispensable emplear el concepto de l\'{\i}mite al punto. Esto no es algo nuevo, dado que siempre que se hace una aproximaci\'on a primer orden o se hace alg\'un c\'alculo aproximado, el f\'{\i}sico siempre emplea el concepto de ``muy peque\~no'' o de infinitesimal. Y aunque estos conceptos son f\'{\i}sicamente muy imprecisos, se logran resultados que est\'an en correspondencia con el experimento, lo cual es otro indicio de la diferenciabilidad en las cantidades f\'{\i}sicas. De otra manera, esto nos indica que toda cantidad f\'{\i}sica tiene una extensi\'on local\footnote{es en este sentido que considero la suavidad de una cantidad}, que en t\'erminos matem\'aticos significa que estas tengan una expansi\'on en serie de Taylor a primer orden. Luego no hay una raz\'on clara y de suficiente peso para abandonar en la f\'{\i}sica las estructuras diferenciales, las cuales son \'utiles y no est\'an en aparente contradicci\'on con nuestra concepci\'on natural, adem\'as de los problemas t\'ecnicos que surgir\'{\i}an al dejarlas a un lado. Por otro lado, el concepto de infinitesimal es algo que ya tiene cupo dentro de las matem\'aticas (el lector bien puede indagar esto, en el an\'alisis no-standard o en la geometr\'{\i}a diferencial sint\'etica), donde el concepto de l\'{\i}mite al punto es reemplazado por c\'alculos algebr\'aicos con n\'umeros infinitesimales. 

Sin embargo, el caracter puntual del \st, aunque forzado por la idea de un espacio-tiempo continuo, si presenta dificultades. Esto debido a que el punto geom\'etrico ``no tiene partes ni magnitud'', lo que significa que su existencia es un concepto no analizable, no resoluble en partes, este solo indica posici\'on (si est\'a o no est\'a) en el \st. Por lo tanto, la pregunta inmediata es si tiene f\'{\i}sicamente sentido esta entidad en nuestra concepci\'on de la {\it natura} (o por lo menos si esto no choca con nuestra concepci\'on del mundo f\'{\i}sico). 

Si tenemos en cuenta que toda medida est\'a asociada a una escala $\ell$ y que por lo tanto toda medici\'on tiene una incertidumbre proporcional a $\ell$, entonces el caracter puntual del \st\ presenta dificultades, ya que no tiene un soporte experimental. Sin embargo, esto no es tan delicado para la actual investigaci\'on te\'orica en f\'{\i}sica cu\'antica, como el hecho de que la medida patr\'on $\ell$ tambi\'en puede presentar una incertidumbre y que esta sea irreducible. Sin embargo, a pesar de que tengamos un l\'{\i}mite a lo f\'{\i}sicamente medible, no implica necesariamente cuantizar el espacio-tiempo o quitarle la estructura suave que le hemos introducido (esto seg\'un los comentarios antes expuestos).

\begin{flushright}
{\it Hacia un principio de incertidumbre irreducible o la imposibilidad del punto geom\'etrico en la f\'{\i}sica.} \bf{(VIII)}
\end{flushright}

Desde un punto de vista f\'{\i}sico, si suponemos la validez de la Relatividad General (RG) a cualquier escala, entonces cualquier sistema f\'{\i}sico o \mpm\ con energ\'{\i}a $E_m$ tiene un radio de Schwarzschild asociado igual a $r_s=2GE_m/c^4$. Por lo tanto, podemos considerar su horizonte de eventos como la frontera del sistema y el interior como su dominio o estructura interna. Si perturbamos esta \mpm\ con un fot\'on, su longitud de onda debe cumplir la condici\'on\footnote{Aqu\'{\i} usaremos los s\'{\i}mbolos $\lessapprox$ ($\gtrapprox$) para indicar ``menor que'' (``mayor que'') y ``del orden de''.}
$$
\lambda\quad =\quad \frac{\hbar c}{E_p}\quad \lessapprox\quad \mbox{di\'ametro}\quad\approxeq\quad 2r_s,
$$
adem\'as $E_m \gtrapprox E_p$, por lo tanto
$$
2r_s\quad\gtrapprox\quad\frac{\hbar c}{E_p}\quad\gtrapprox\quad\frac{\hbar c}{E_m}=\frac{\hbar c}{r_sc^4/2G},\qquad\Rightarrow\qquad r_s\quad\gtrapprox\quad\sqrt{\frac{G\hbar}{c^3}}.\nonumber
$$
Esto implica que no es adecuado considerar los sistemas f\'{\i}sicos como puntuales. 

Por otro lado, los conceptos combinados de RG y de MC requieren una situaci\'on en la cual ambas condiciones deben ser satisfechas simultaneamente. Luego, el m\'{\i}nimo radio para un objeto elemental ser\'{\i}a su radio de Schwarzschild ($r_s=2Gm_o/c^2$), y el m\'aximo tama\~no su longitud de Comptom ($\lambda_c=\hbar/m_oc$), entonces exijimos que $2r_s\lessapprox\lambda_c$. Esto implica que 
$$
\frac{\hbar}{2m_o c}\quad\gtrapprox\quad r_s\quad\Rightarrow\quad 
\sqrt{\frac{G\hbar}{c^3}}\quad\gtrapprox\quad r_s,
$$
luego 
$$
r_s\quad\approxeq\quad\sqrt{\frac{G\hbar}{c^3}}.
$$
Esto ya nos permite hacer una estimativa de la m\'{\i}nima distancia medible $\delta l$ entre dos part\'{\i}culas libres, usando el principio de incertidumbre de Heisenberg $\Delta x \Delta p \geq \hbar/2$, con $\Delta p \approxeq m_oc$ obtenemos que 
$$
\delta l\quad\geq\quad\Delta x\quad\geq\quad\frac{\hbar}{2\Delta p}\ =\ \frac{\hbar}{2m_o c}\quad \gtrapprox\quad r_s. 
$$
Por lo tanto $\delta l$ tiene un l\'{\i}mite irreducible
\begin{equation}
\delta l\quad\geq\quad\sqrt{\frac{G\hbar}{c^3}}={\cal L}_p\quad\simeq\quad 1.61*10^{-35} metros. 
\label{l-irred}
\end{equation}
De manera similar el lector puede comprobar que tambi\'en tenemos una m\'{\i}nima cantidad de tiempo medible $\delta t$, con 
\begin{equation}
\delta t\quad\geq\quad\sqrt{\frac{G\hbar}{c^5}}={\cal T}_p\quad\simeq\quad 5.39*10^{-44} segundos. \label{t-irred}
\end{equation}
Esto, no solo nos indica la posibilidad de una incertidumbre en el \st\ y en la medici\'on, sin\'o tambi\'en la imposibilidad de obtener una medida f\'{\i}sica exacta, de la misma manera que en termodin\'amica es imposible obtener el cero absoluto en una medici\'on de temperatura de un sistema estad\'{\i}stico (esto sin importar la escala). 

Luego, cualquier intervalo de espacio o de tiempo m\'as peque\~no que estos no puede ser medible. Por lo tanto, los conceptos de punto y de instante para representar el lugar espacio-temporal de una \mpm\ en el \st\ dejan de tener sentido, ya que siempre existir\'a una incertidumbre experimental y su l\'{\i}mite inferior ser\'a algo inherente al sistema y no una deficiencia del instrumento. Adem\'as es posible que exista una regi\'on de indistinguibilidad de la \mpm, en la cual dos puntos diferentes son indistinguibles, y que es menor a la m\'{\i}nima longitud medible. 

\begin{flushright}
{\it Los reales y la representaci\'on puntual conlleva a infinitos.} \bf{(IX)}
\end{flushright}

Por otro lado, en la actual investigaci\'on te\'orica en f\'{\i}sica, uno de los problemas que m\'as dificultad ha presentado es el problema de las singularidades, el cual est\'a asociado con la interpretaci\'on de part\'{\i}cula puntual en el \st. Por esto, es que surgen infinitos en las ecuaciones, lo cual aparece con mayor severidad en la Teor\'{\i}a Cu\'antica de Campos (TCC) (por ejemplo cuando se calcula la auto-energ\'{\i}a electromagn\'etica de un electr\'on acotado). Aunque este problema se ha aminorado por subsiguientes mejoras en la teor\'{\i}a, el problema de infinitos permanece aun con nosotros, en particular cuando se quiere introducir la interacci\'on gravitacional en las ecuaciones de campo, luego esquem\'aticamente tenemos que 
$$
\hbox{``\sf Reales + Representaci\'on puntual}\to \infty \hbox{ ''}
$$

Es por esto que, nuevas representaciones del \st\ y de \mpm\ han surgido con el prop\'osito de obtener una t. cu\'antica que incluya la gravedad y sin infinitos:
\begin{itemize}
\item La teor\'{\i}a de cuerdas: la cual considera la \mpm\ como una cuerda o d-brana, es decir una vibraci\'on o excitaci\'on en una variedad $\Im$ inmersa sobre alg\'un $\nr^n$ con $n>4$, de tal manera que las dimensiones extras son consideradas como par\'ametros internos o ``variables ocultas'' que solo muestran su existencia en el orden de la escala de Planck ($10^{-32}$ metros) o m\'as all\'a. Sin embargo, un \st\ con $dim>4$ no lo podemos percibir, entonces solo es posible una ``medici\'on'' indirecta de las dimensiones extras. El \'unico referente de esta representaci\'on ser\'{\i}a el universo f\'{\i}sico $4$-$dim$ que percibimos, y el l\'{\i}mite efectivo ($\Im\downarrow \nr^4$) de esta teor\'{\i}a a 4 dimensiones aun no se ha logrado.

\item Espacios reticulares: aqu\'{\i} se considera un \st\ descrito por un ret\'{\i}culo parcialmente ordenado\footnote{Donde la cardinalidad de esta estructura estar\'{\i}a en correspondencia con la cardinalidad de $\nn$} y con una longitud fundamental $L$, el cual reproducir\'{\i}a una f\'{\i}sica sobre $\nr$ en el $\lim L\to 0$. Luego, una TCC con esta estructura solo trabajar\'{\i}a a distancias cuantizadas y mayores que $L$, y por lo tanto todas las integrales divergentes podr\'{\i}an ser removidas a distancias menores o del orden de magnitud de $L$ (o momento $\hbar/L$). Pero esta estructura presenta conflitos con los teoremas de conservaci\'on y los difeomorfismos de invarianza en el vacio. Por otro lado $L$ se podr\'{\i}a contraer (o dilatar) en el l\'{\i}mite de $v\to c$ ($v\to 0$), entonces que valor darle a $L$?. Adem\'as, qu\'e sentido tendr\'{\i}a aqu\'{\i} $\pi$ u otros n\'umeros Irracionales. 

\item Espacios cu\'anticos: este esquema extiende por {\sf analog\'{\i}a} la no-conmutatividad de ($X,P$) a la estructura del \st. Por lo tanto, as\'{\i} como es imposible medir con exactitud el momento y la posici\'on (dadas las relaciones de Heisenberg), de la misma forma hay una imposibilidad en tener un valor exacto o puntual en el \st, entonces
$$
\frac{[X_i,P_i]=i\hbar}{\Delta X_i\Delta P_i\geq \frac{1}{2}\hbar} 
\qquad\sim\qquad 
\frac{[X_i,X_j]=\sigma_{ij}}{\Delta X_i\Delta X_j\geq \frac{1}{2}\sigma_{ij}}.
$$
Sin embargo, esta estructura posibilita te\'oricamente hacer una medici\'on exacta, si es que esto tiene alg\'un sentido f\'{\i}sico.
\end{itemize}

Estas soluciones, al igual que cualquier otra soluci\'on propuesta hasta ahora, de alguna u otra forma representan el \st\ como una variedad o una estructura inmersa sobre alg\'un producto cartesiano de $\nr$, y no se considera en duda que la f\'{\i}sica en su versi\'on final (es decir al tomar alg\'un l\'{\i}mite) debe estar descrita por n\'umeros reales. Sin embargo, nadie cuestiona si los n\'umeros reales tal como los conocemos, son la estructura matem\'atica adecuada para representar el \st\ y la \mpm. De tal forma que unifique f\'{\i}sica cu\'antica $\&$ gravedad, o l\'ogica $\&$ geometr\'{\i}a. 

Por un lado $\nr$ posee todas las buenas propiedades matem\'aticas que a uno le gustar\'{\i}a tener, pero son estas, cualidades f\'{\i}sicas?. Hasta qu\'e circunstancia este objeto representa todo lo que consideramos f\'{\i}sico?, son los reales necesarios y suficientes?. Por un lado, sabemos que $\nr$ es m\'as de lo que la f\'{\i}sica experimental necesita y tampoco son suficientes, si hay una imposibilidad en tener estructuras puntuales, de la misma manera que hay una imposibilidad de que nuestros sentidos puedan percibir el punto y el instante. Adem\'as $\nr$ tiene la propiedad de {\sf buen orden}\footnote{\\
Definici\'on: $A$ es una {\sf clase parcialmente ordenada} (CPO), si este posee una relaci\'on (de orden) que es: reflexiva ($\forall x,x\leq x$), antisim\'etrica (si $x\leq y, y\leq x\Rightarrow x=y$) y transitiva (si $x\leq y,y\leq z\Rightarrow x\leq z$).\\
Definici\'on: Si $A$ es CPO, $A$ tiene {\sf buen orden} si cualquier subclase no vac\'{\i}a de $A$ tiene un elemento menor.\\
Definici\'on: Si $A$ es CPO, $A$ es {\sf completamente ordenado} si cualquier dos elementos son comparables, es decir si puedo establecer una relaci\'on de orden entre cualquier dos elementos ($\forall x,y\in A$, $x\leq y$ o $x\geq y$).\\
Teorema: si $A$ tiene {\sf buen orden} $\Rightarrow$ $A$ es {\sf completamente ordenado}. Por ejemplo $\nn,\nq,\nr$ tienen la propiedad de buen orden y por lo tanto son conjuntos completamente ordenados.}, lo cual no siempre es posible de contrastar ya que toda medida f\'{\i}sica tiene una incertidumbre, y con mayor raz\'on si esta es irreducible.

El problema es bastante delicado, porque toda la estructura matem\'atica de la f\'{\i}sica (e incluso de las matem\'aticas standard) est\'a construida en una teor\'{\i}a de conjuntos que hace posible la existencia de conjuntos s\'{\i}ngleton $\{x\}$ (o subconjuntos de un solo elemento), lo cual solo es posible, gracias al {\sf Axioma de Escojencia} (\AE)\footnote{\it Sea $X$ una colecci\'on de conjuntos no-vacios. Entonces nosotros podemos escojer un miembro de cada conjunto en esta colecci\'on. Indicado m\'as formalmente, existe una funci\'on $f$ definida sobre $X$ tal que para cada $S\in X$, $f(S)$ es un elemento de $S$.}, el cual est\'a en equivalencia con la propiedad de {\sf buen orden} en la teor\'{\i}a de conjuntos. Este axioma nos indica que existe alguna funci\'on $f$ que puede escojer un elemento de cada conjunto en la colecci\'on, pero no puede indicarnos como esta funci\'on puede ser definida, simplemente dice que $f$ existe. Y como todas las pruebas que involucran el {\AE}, son solo pruebas de existencia y son siempre no-constructivas\footnote{Como una muestra, estos son algunos teoremas centrales en varias \'areas de las matem\'aticas que requieren el {\AE} (o una de sus versiones d\'ebiles): Cualquier espacio vectorial tiene una base; El teorema de Hahn-Banach en el an\'alisis funcional, seguido de la extensi\'on a funcionales lineales; El teorema de Banach-Alaoglu sobre compacidad de conjuntos de funcionales; Un espacio uniforme es compacto si y solo si este es completo y totalmente acotado; Cualquier espacio de Tychonoff tiene una compactificaci\'on de Stone-Cech; El teorema de representaci\'on de Stone, el cual dice que cualquier \'algebra boleana es isomorfa a alg\'un \'algebra boleana de conjuntos; etc...}. 

Visto de otra manera, este axioma hace posible definir una cantidad f\'{\i}sica sobre un punto del \st, o definir el concepto de funci\'on como una relaci\'on entre puntos. Luego, no tiene sentido f\'{\i}sico hablar de objetos que est\'an basados en el {\AE} (o en la idea de punto)\footnote{Por ejemplo, hablar de part\'{\i}cula puntual, de instante, velocidad instantanea, part\'{\i}culas puntuales, hablar de dimensiones a una escala inferior a la m\'{\i}nima medible, hablar de estructuras Hausdorf, funciones delta de Dirac, l\'{\i}mite $\to 0$, l\'{\i}mite $\to \infty$, dividir un intervalo infinitamente, etc.}, si NO hay una manera de construirlos, solo se supone su existencia, NO es posible contrastarlos con el experimento, y CHOCAN con la idea de una incertidumbre m\'{\i}nima e irreducible?. 

\begin{flushright}
{\it Abstracci\'on $\&$ experimentaci\'on.} \bf{(X)}
\end{flushright}

Desde hace algo m\'as de un siglo, en los c\'{\i}rculos matem\'aticos se ha llevado una fuerte discusi\'on y competencia sobre la manera de construir el conocimiento matem\'atico. Por un lado, los matem\'aticos intuisionistas piensan que no hay validez en considerar objetos matem\'aticos donde solo se infiere su existencia, y que por lo tanto ning\'un supuesto puede basarse en la falsedad de su negaci\'on. Estos dicen que nuestra mente solo puede darle cavida a algo finito, es decir solo puedo contar un n\'umero finito de cosas, y niegan la posibilidad del infinito entre otras cosas, porque su existencia solo se puede entender por lo que no es, es decir por lo que no es finito. Estos tambi\'en niegan la posibilidad del punto geom\'etrico como un elemento del continuo real, porque nuestros sentidos no lo pueden captar, y por consiguiente el {\AE} no tiene tampoco validez all\'{\i}. Para eludir las supuestas contradicciones (paradoja de Cantor, paradoja de Russell, paradoja sem\'antica de Berry, etc..), los matem\'aticos intuisionistas suponen que no necesariamente sea v\'alido el {\sf Axioma de Tercio Excluso} ({\sf ATE})\footnote{Para cualquier proposici\'on $P$, se tiene en un tiempo solo uno de estos: $P$ es $\top$ o $\neg P$ es $\top$}, lo que implica que los conjuntos cuciales en las paradojas ya no pueden ser construidos y los argumentos esenciales no pueden ser examinados. Por un lado se gana en estructura y las matem\'aticas se vuelven m\'as ricas en lenguaje, pero por otro lado se pierde mucho del conocimiento matem\'atico acumulado, ya que algo solo es verdadero si puedo mostrarlo o construirlo. Por otro lado, en las matem\'aticas standard, se considera una estructura matem\'atica bivalente, en la cual s\'{\i} se puede hacer una demostraci\'on usando el {\sf ATE}, y tiene sentido considerar objetos donde solo se postula su existencia. Los matem\'aticos conjuntistas abogan que s\'{\i} es posible deshacerse de las paradojas sin recurrir a otras l\'ogicas. 

Es aqu\'{\i}, donde se debe reconocer que el conocimiento matem\'atico es un juego de coherencia, donde la l\'ogica antecede y es incidental a las matem\'aticas. Por lo tanto, es posible colocar los supuestos o reglas de deducci\'on formal que se quiera, siempre y cuando se mantenga consistencia y se diga algo\footnote{Tenga en cuenta que si algo puede ser falso y verdadero al tiempo, o algo y su negaci\'on son verdaderas, o todos los s\'{\i}mbolos de un lenguaje son el mismo, se tiene entonces que: todo puede ser falso y verdadero, o todo puede ser cierto, o que se ha dicho nada.}. El juego matem\'atico es un juego de la mente humana y no necesita tener un referente con nuestra realidad o con nuestra percepci\'on, y aunque no podamos percibir un punto o un instante de tiempo, esto no significa que nuestra mente no lo pueda considerar, como de hecho lo hacemos; hablamos de puntos con plena seguridad y comodidad. Luego, carece de sentido poner a competir esquemas cuando de hecho ni siquiera son comparables, son simplemente diferentes.

Si bien, nuestra mente matem\'atica tiene plena libertad, en considerar solo de un objeto su existencia, ya hemos abierto una brecha entre lo que es matem\'aticamente {\sf coherente} y lo que es f\'{\i}sicamente perceptible, entre la abstracci\'on y el experimento. Dado que hasta ahora no habiamos tenido una contradicci\'on entre nuestro modelo y nuestra concepci\'on {\it natural}. Y aunque la existencia de un objeto en el lenguaje matem\'atico es cuesti\'on de {\sf coherencia}, en la f\'{\i}sica esto no es suficiente, adem\'as es necesario que sus inferencias y atributos representen y est\'en en correspondencia con nuestra percepci\'on f\'{\i}sica (es decir que sus consecuencias y propiedades sean cohe-rentes con nuestra {\it realidad}). De tal forma que, el significado y las propiedades de un objeto solo est\'en bien definidos cuando este sea conjugado con otros objetos de la estructura contemplada. Por lo tanto, exijimos el siguiente principio, que llamaremos para efectos pr\'acticos: {\sf Principio de Acci\'on Constructiva} ({\sf PAC}),
\begin{quote}
En un buen modelo de la {\it realidad}, la existencia de un objeto matem\'atico debe ser equivalente a la posibilidad de su {\sf construcci\'on expl\'{\i}cita} y a la posibilidad de su {\sf contrastaci\'on f\'{\i}sica}. De tal forma que sus atributos e inferencias f\'{\i}sicas sean {\sf perceptibles}, y donde la refutaci\'on de no-existencia no significa necesariamente que es posible encontrar una prueba de existencia. 
\end{quote}

Por un lado, este principio nos quita la posibilidad de tener objetos puntuales, los cuales no son contrastables y est\'an en contradicci\'on con una incertidumbre irreducible. Pero, por otro lado se est\'a exijiendo a nuestra estructura matem\'atica que, lo que no es {\sf contrastable} y/o no {\sf perceptible} (o no puede serlo), no es f\'{\i}sico y por lo tanto debe estar fuera de las consideraciones y modelos f\'{\i}sicos.\footnote{Para efectos pr\'acticos definiremos {\sf construible, perceptible} y {\sf contrastable} como sigue; se dice que un objeto es {\sf construible} si es posible mostrar o construir este objeto expl\'{\i}citamente apartir de supuestos elementales e intuitivos (aqu\'{\i} considero que algo no es intuitivo, si de \'el solo puedo considerar la posibilidad de su existencia). Un objeto de nuestra mente es {\sf contrastable} si existe una manera perceptible de evidenciar este objeto. Y se cree que algo es {\sf perceptible}, si nuestra mente y nuestros sentidos (como un colectivo) nos pueden mostrar y dejar una impresi\'on (ya sea con o sin ayuda de un instrumento o trasductor).}

\begin{flushright}
{\it Una necesaria evoluci\'on a modelos constructivistas.} \bf{(XI)}
\end{flushright}

Al atacar la idea puntual en los modelos f\'{\i}sicos, y encontrar que hay una imposibilidad de tener este tipo de estructuras en la f\'{\i}sica, nos hemos visto obligados a adaptar y modificar el lenguaje matem\'atico, para esto hemos formulado un principio que nos exije que los objetos matem\'aticos de la f\'{\i}sica deban ser {\sf construibles, contrastables} y {\sf perceptibles}. Lo que implica que es necesario cambiar nuestra estructura l\'ogica, dado que ning\'un supuesto e inferencia puede estar basado en una prueba de existencia, como en algunos casos se hace. Luego, no aceptamos necesariamente el {\sf ATE}, dentro de nuestra ``l\'ogica f\'{\i}sica'' ya que este permite objetos no construibles. Por consiguiente, no tenemos el {\AE} en nuestros axiomas matem\'aticos, y por lo tanto no hay estructuras puntuales\footnote{De otra manera, el lector puede darse que cuenta que es realmente la ley de tercio excluso de la l\'ogica, lo que hace posible construir funciones tipo delta de Dirac.}.

De otra manera, aqu\'{\i} estamos privilegiando una visi\'on intuisionista para la f\'{\i}sica. Una visi\'on constructivista, en la cual, cualquier afirmaci\'on solo puede ser mostrada de manera contructiva, o explicitamente apartir de supuestos que sean intuitivos y perceptibles. Luego, solo es posible mostrar una afirmaci\'on por negaci\'on, si nuestros objetos de estudio solo ofrecen dos opciones o estados posibles\footnote{por ej. si una pantalla muestra las im\'agenes con solo ``blanco $\&$ negro'', se tiene que: ``si un p\'{\i}xel no es blanco $\Rightarrow$ este p\'{\i}xel es negro.''}. Y se muestra la falsedad de una afirmaci\'on, si esta afirmaci\'on implica o conlleva a una contradicci\'on en nuestra estructura. Aqu\'{\i} ya se puede notar, que el lenguaje es m\'as rico en estructuras, dado que al no tener tercio excluso, implica entre otras cosas que, nuestra l\'ogica es ahora una l\'ogica polivalente. En donde, el objeto que representa los valores de verdad $\Omega$, contiene m\'as valores que $\{\top,\perp\}$, adem\'as de que no necesariamente hay una relaci\'on de orden entre todos los valores de verdad (es decir puede haber dos valores para los cuales no se puede saber cual es m\'as cierto). Por consiguiente, el \'algebra de esta l\'ogica es un \'algebra de Heyting y no un \'algebra Boleana.

Para reconstruir nuestro modelo matem\'atico de la f\'{\i}sica, no usaremos los conceptos de la teor\'{\i}a de conjuntos, en vez de esto nos aprovecharemos de la teor\'{\i}a de categor\'{\i}as\footnote{una categor\'{\i}a consiste de objetos (simbolizados con letras may\'usculas), y de flechas (etiquetadas por letras minusculas), adem\'as de la igualdad (=). De tal manera que toda flecha tiene asociada dos objetos que se llaman $Dominio$ u objeto de partida y $Codominio$ u objeto de llegada, con los siguientes axiomas: 
\begin{itemize}
\item Composici\'on: si $f:A\to B$, $g:B\to C$, entonces existe $h:A \to C$ tal que $h=g\circ f$,
\item Asociatividad: si $f:A\to B$, $g:B\to C$ y $h:C\to D \Rightarrow h\circ (g\circ f)=(h\circ g)\circ f$, y
\item Identidad: para cada objecto $X$ existe una flecha $id_X : X \to X$ llamada la identidad para $X$, tal que para cualquier flecha $f:A \to B$ tenemos $id_B \circ f = f \circ id_A$ 
\end{itemize}}, porque esta captura en buena parte la esencia de nuestra aproximaci\'on\footnote{Donde los objetos pueden indicar nuestros objetos con ``sentido f\'{\i}sico'' y las flechas entre objetos como las relaciones entre los objetos, las cuales definir\'an las propiedades e inferencias f\'{\i}sicas}. En categor\'{\i}as, podemos eludir los problemas que genera una teor\'{\i}a conjuntista (en donde el concepto de funci\'on es reducido al concepto de conjunto de puntos), debido a que las transformaciones (flechas) entre estructuras (objetos), juegan un papel aut\'onomo y no subordinado entre si. Por consiguiente, la noci\'on de transformaci\'on puede admitir una interpretaci\'on, en la cual una cantidad variable depende funcionalmente de otra, pero la correspondiente funci\'on no se describe como un conjunto (de pares ordenados) de puntos. Adem\'as de que en categor\'{\i}as desaparece el concepto de verdad absoluta, y en su lugar aparece el concepto de invariancia, el cual es v\'alido en todos los marcos locales de trabajo. 

Dado que cualquier cantidad medible se puede aproximar por un \'unico n\'umero real, se tiene entonces que $\nr$ es el objeto fundamental de nuestra descripci\'on matem\'atica de la realidad, el cual se define como el conjunto de partes de $\nn$ o $\nr=2^\nn$. Luego, cualquier n\'umero se puede representar como una cadena de $\{0,1\}$, o se puede encontrar haciendo solo preguntas de falso y verdadero, es decir $\nr=\Omega^\nr$ con $\Omega=\{\top,\perp\}$. Pero al trabajar en una l\'ogica constructivista, este objeto es afectado dr\'asticamente, ya que $\Omega$ es ahora un objeto m\'as complejo. Este posee una valoraci\'on polivalente, por consiguiente $\nr$ toma una forma m\'as rica en estructuras, donde no necesariamente es posible encontrar un n\'umero con solo preguntas de falso y verdadero, y podriamos encontrarnos con el caso en donde ni siquiera un n\'umero es etiquetable. Por lo tanto surge una nueva caracter\'{\i}stica, y es que cada elemento o ``punto'' $x$ de un objeto, puede tener una vecindad de ``puntos'' que le son inseparables, lo cual conlleva al concepto de {\sf m\'onada} de $x$ como ``el objeto de puntos de indistinguibilidad'', y que notaremos como $[x]$. 

En cierta forma todo punto tiene una nube borrosa, en la cual puede haber 2 elementos que son ``distintos'' pero indistinguibles. Como un ejemplo f\'{\i}sico tomese la siguiente sucesi\'on de puntos separados por una distancia de $1$ cm. $:\frac{1}{2}$ cm. $:\frac{1}{4}$ cm. $:\frac{1}{8}$ cm. $:\frac{1}{16}$ cm. $:\frac{1}{32}$ cm.$:\frac{1}{64}$ cm.
$$
\dot\uparrow\hspace{8.05 mm}
\dot\uparrow\quad\todu\quad\dot\uparrow\hspace{3.05 mm}
\dot\uparrow\quad\todu\quad\dot\uparrow\hspace{0.55 mm} 
\dot\uparrow\quad\todu\quad\dot\uparrow\hspace{-0.7 mm}
\dot\uparrow\quad\todu\quad\dot\uparrow\hspace{-1.325 mm} 
\dot\uparrow\quad\todu\quad\dot\uparrow\hspace{-1.6375 mm}
\dot\uparrow\quad\todu\quad\dot\uparrow
$$
El lector puede notar que en la sexta partici\'on de la unidad ya hay una indistinguibilidad entre los puntos, a pesar de que te\'oricamente la distancia entre estos es de $1/2^6$ cm. Esto ocurre en parte por la resoluci\'on gr\'afica de este documento, pero realmente se debe a que estos son puntos f\'{\i}sicos, y por lo tanto tienen un tama\~no (de lo contrario usted no podr\'{\i}a medir algo). Entonces, en una divisi\'on finita de la unidad, uno de estos puntos entra en el lugar de ocupaci\'on del otro, o en el conjunto de indistinguibilidad mutua o m\'onada (que depende del tama\~no de los puntos, es decir depende de su ``di\'ametro''), lo cual no significa que estos sean el mismo punto. Luego f\'{\i}sicamente carece sentido decir que la distancia entre estos puntos es $\simeq 0,015$ cm, porque realmente la distancia mutua es nula o cero. Por consiguiente, los ``puntos'' ahora tienen estructura interna, sin que esta tenga una medida no nula. Como una ilustraci\'on formal y pedag\'ogica, adaptaremos las estructuras mon\'adicas a un esquema de medici\'on standard. Donde $\Delta$ representa el objeto de las mediciones f\'{\i}sicas, $\ell$ una escala de medici\'on y $\Delta/\ell$ tiene la propiedad de ser un campo algebraico (ver aparte No V),
{\definition Conjunto de medici\'on nula de orden $\ell/n$}
$$
[\hbox{\O}]_{\ell/n}=\{x\in \Delta/\ell| \mbox{ para alg\'un } n,\ |x|\ell\lesssim\ell/n\}.
$$
De otra manera, dada una escala de medici\'on $\ell$, se tiene que $\ell/n$ es la resoluci\'on m\'as peque\~na definible, luego cualquier medici\'on por debajo de este valor, es nula.\footnote{Aqu\'{\i} usaremos los s\'{\i}mbolos $\lesssim$ ($\gtrsim$) para indicar ``menor que'' (``mayor que'') o ``del orden de''.}
{\theorem Equivalencia de orden ${\ell/n}$: $(x\sim y)_{\ell/n}$ si $x-y\in[\hbox{\O}]$. }
{\proof 
\begin{itemize}
\item $(x\sim x)_{\ell/n}$; $\quad |(x-x)|\ell=0\ell\lesssim\frac{\ell}{n}$,
\item $(x\sim y)_{\ell/n}\Rightarrow (y\sim x)_{\ell/n}$; $\quad |(x-y)|\ell=|(y-x)|\ell\lesssim\frac{\ell}{n}$,
\item $(x\sim y)_{\ell/n}$, $(y\sim z)_{\ell/n'}\Rightarrow (x\sim z)_{\ell/k}$; 
$\quad |x-y+y-z|\ell\leq |x-y|\ell+|y-z|\ell\lesssim\frac{\ell}{n}+\frac{\ell}{n'}\lesssim\frac{\ell}{m/2}\lesssim\frac{\ell}{[\![m/2]\!]}$, siendo $m=min\{n,n'\}$, y $[\![m/2]\!]$: {\it parte entera de $m/2$}.
\end{itemize}}
{\theorem Sea $n\geq m$, luego $(x\sim y)_{\ell/n}\Rightarrow (x\sim y)_{\ell/m}$, o de otra manera $[x]_{\ell/n}\subseteq [x]_{\ell/m}$. }{\proof $|(x-y)|\ell\lesssim\frac{\ell}{n}\leq\frac{\ell}{m}$ .}
{\definition Se define la m\'onada para una medici\'on $x$ como} 
$$
[x]_{\ell/n}=\{x'\in \Delta/\ell| (x'\sim x)_{\ell/n}\}.
$$
Luego, dos puntos pertenecen a la misma m\'onada, si estos tienen la misma medida, o si la diferencia relativa entre estos es menor que la resoluci\'on de medida.
{\theorem $[x]_{\ell/n}=[y]_{\ell/n}$ si y solo si se cumple alguno de estos $(x\sim y)_{\ell/n}$ o $(x\in[y]_{\ell/n})$ o $(y\in[x]_{\ell/n})$. }{\proof Obvio.}
{\theorem $[0]_{\ell/n}=[\hbox{\O}]_{\ell/n}$}. {\proof $0\in [\hbox{\O}]_{\ell/n}$.}
{\theorem $[0+x]_{\ell/n}=[x]_{\ell/n}$}. {\proof $(0+x\sim x)_{\ell/n}$.}

Por consiguiente, cualquier medici\'on de la forma $x\ell\pm \ell/2n$ puede ser representada en el anterior esquema como $[x]_{\ell/n}\ell$, el cual nos indica el conjunto de nulidad o de indistinguibilidad. Pero si toda medida f\'{\i}sica tiene una incertidumbre irreducible $\delta\ell={\cal L}_p$ (ver aparte No VIII), entonces carece de sentido hablar de mediciones por debajo (o del orden) de esta cota, dado que cualquier medici\'on menor (o del orden) de $\delta\ell$ es una medici\'on nula. Luego toda medici\'on tiene un conjunto de indistinguibilidad o una m\'onada $[x]_{\delta\ell}$, en cierta forma el \st\ tambi\'en tiene una resoluci\'on, as\'{\i} como este documento tiene una resoluci\'on gr\'afica. Ahora bien, si hacemos una medici\'on f\'{\i}sica y comenzamos a mejorar nuestra precisi\'on o hacemos nuestra escala $\ell/n$ m\'as peque\~na, nos vamos a topar con la condici\'on experimental l\'{\i}mite, de que la distancia m\'as peque\~na posible entre dos objetos f\'{\i}sicos es $0$ y no $\delta\ell={\cal L}_p\neq 0$ (tenga en cuenta el ejemplo anterior, en el que se considera la distancias entre dos puntos). De otro modo, nadie puede decir que ha medido una distancia ${\cal L}_p$, ya que no puede tener un patr\'on escala con que comparar y que pueda medir una distancia $\delta\ell={\cal L}_p$. Por consiguiente, {\it $\delta\ell$ no es decidible (o no resoluble); es decir no es cierto que $\delta\ell=0$ y tampoco que $\delta\ell\neq 0$.} Teniendo en cuenta que esta formulaci\'on es ya radicalmente diferente a los principios matem\'aticos que hasta ahora han permitido describir la f\'{\i}sica, y en donde el concepto de punto es reemplazado por el concepto de m\'onada, llamaremos a esto {\sf F\'{\i}sica Mon\'adica}, y que notaremos como \MP. 
\begin{flushright}
{\bf F\'{\i}sica Mon\'adica \MP}
\end{flushright}

Ahora nos enfocaremos en el observable por excelencia: El \ST, y a pesar de que hay mucha discusi\'on en cuanto a su definici\'on, s\'{\i} hay concenso en las propiedades que lo describen, es decir por las propiedades de su medici\'on. Luego, el primer objeto que suponemos en \MP, es el objeto que nos ayudar\'a a definir el \st, y que notaremos como $\kk$. Al cual le asociamos los siguientes postulados:

{\axiom: $\kk$ posee dos elementos $0,1$ y las siguientes relaciones $-:\kk\to\kk$, $+:\kk\times\kk\to\kk$, $*:\kk\times\kk\to\kk$, tal que para cualesquier variables $x,y,z$ en $\kk$, tenemos que:\label{axiom1}}
$$
1*x=x,\quad (x*y)*z=x*(y*z),\quad\hbox{si }x\neq 0,(\exists y\ |\ x*y=1)\wedge (\forall y\ |\ x*y=y*x),
$$
$$0+x=x,\quad (x+y)+z=x+(y+z),\quad x+(-x)=0,\quad x+y=y+x,
$$
$$x*(y+z)=(x*y)+(x*z)
$$
Adem\'as de que\footnote{Aqu\'{\i} ``$\mathbb{M\hspace{-1.2 mm}P}\vdash \Phi$'' se lee como ``En $\mathbb{M\hspace{-1.2 mm}P}$ es cierto $\Phi$'', y ``$\neg \Phi$'' como ``No $\Phi$''}
{\axiom $\MPm\vdash\ \neg(0=1)$.\label{axiom2}}\\
No es dificil mostrar que:
{\theorem $\MPm \vdash\ (\forall x), 0*x=x*0=0$\label{T1}}\\
{\proof $0x=(a-a)x=ax-ax=0=xa-xa=x(a-a)=x0$.}

Supuesta una relaci\'on de equivalencia $\sim$, se define:
{\definition objeto nulo o de nulidad}
$$
[\hbox{\O}]=\{x\in {\cal K}\quad|\quad x\sim 0\}.
$$
{\definition Se define la m\'onada para un elemento $x$ como} 
$$
[x]=\{x'\in {\cal K}\quad|\quad x'-x \in [\hbox{\O}]\}.
$$
Luego,
{\theorem $[\hbox{\O}]=[0]$.} {\proof $0\in[\hbox{\O}]$.}
{\theorem $\MPm \vdash\ \kk$ contiene el objeto de los n\'umeros racionales}\\
{\proof Dado que todo Topos contiene el objeto de los n\'umeros naturales, entonces $\kk$ contiene a los enteros, los cuales forman un anillo conmutativo con unidad. Por consiguiente, para cualquier $n\in\nz$ con $n\neq 0$, este tiene un inverso mutiplicativo. Luego $\kk$ contiene a los n\'umeros racionales. Si ${\kk}$ ``contiene'' a $\ni$, se tiene que ${\kk}$ ``contiene'' a $\nr$, lo que nos permite obtener el continuo de la recta, por consiguiente notaremos a ${\kk}$ como $\Re$ cuando se quiera indicar esto. 

\begin{flushright}
{\bf Modelos para las estructuras mon\'adicas}
\end{flushright}

Debido a que el anterior esquema es muy general, es necesario plantear modelos para las estructuras mon\'adicas. Entonces nos aprovecharemos de los objetos nilpotentes de la geometr\'{\i}a diferencial sint\'etica (GDS), aunque esto no es exactamente GDS, como el lector podr\'a notar m\'as adelante:
{\definition Se define la relaci\'on de equivalencia de orden $n$ como} 
$$
(\mu \sim_n 0 \quad \Leftrightarrow\quad \mu^n=0) \qquad \mbox{ con }\quad n\in \nn\quad\mbox{ y }\quad n\geq 1.
$$
Luego, el objeto de nulidad definido en la anterior subsecci\'on, toma la forma\footnote{Siempre que queramos representar un objeto lo haremos con letras may\'usculas y un n\'umero con la propiedad nilpotente lo notaremos con letras griegas.}
$$
[\hbox{\O}]_n=\{\mu\in {\kk}\quad|\quad \mu^n=0\}.
$$
Para $n=1$ se puede notar que $[\hbox{\O}]_n$ es el s\'{\i}ngleton $\{0\}$. La m\'onada para $x$ toma la forma
$$
[x]_n=\{x'\in {\kk}\quad|\quad x'-x\in [\hbox{\O}]_n\}.
$$
De otra manera se tiene que $x'\in[x]_n$, si $x'=x+\mu$ con $\mu^n=0$. Luego
{\theorem $[0]_n=[\hbox{\O}]_n$, adem\'as $[x]_n=x+[0]_n=x+[\hbox{\O}]_n$.}\\
{\proof $0\in [\hbox{\O}]_n$. Si $\mu\in [\hbox{\O}]_n$, entonces $x+\mu\in[x]_n$, y si $x'\in [x]_n \Rightarrow x'=x+\mu$.}\\
Ahora construiremos el siguiente axioma para darle forma a la anterior estructura\footnote{Aqu\'{\i} $A^B$ representa el objeto de las transformaciones de $B\to A$.} 
{\axiom $\alpha:\kk^n\to\kk^{[\hbox{\O}]_n}$ es un isomorfismo.\label{axiom3}}\\
Siendo $\kk^n$ n-productos cartesianos de $\kk$. Esto nos permite tener entre otras cosas: el principio de cancelaci\'on m\'{\i}nima,
{\theorem $\MPm\vdash(\forall w\in\kk)$ $((w*x=w*y)\Rightarrow (x=y))$.}\\
{\proof Inmediato del anterior axioma.}
{\theorem Si $n>1$, $[\hbox{\O}]_n\neq\{0\}$, de otra forma $\MPm \vdash\neg(\forall\mu\in[\hbox{\O}]_n)(\mu=0)$.}\\
{\proof Si $(\mu=0)\Rightarrow (\mu*0=\mu=\mu*1)\Rightarrow 0=1$. Pero el axioma (\ref{axiom2}) nos indica que $0\neq 1$, luego $\neg(\mu=0)$\footnote{Siempre que se use $\mu \neq 0$, se est\'a indicando $\neg(\mu=0)$.}.
{\theorem $[\hbox{\O}]_n$ es no decidible, es decir $\MPm\vdash\neg(\forall \mu\in[\hbox{\O}]_n)(\mu=0\vee \mu\neq 0)$.}\\
{\proof Si $\mu\neq 0\Rightarrow \exists \nu\ |\ \mu*\nu=1$, luego $1=(\mu*\nu)^n=0$. Pero el axioma (\ref{axiom2}) nos indica que $0\neq 1$, por consiguiente $\neg(\mu\neq 0)$\footnote{El lector puede notar que aqu\'{\i} $\neg\neg (\mu=0)$ no es lo mismo que ($\mu=0$)}. Por otro lado el anterior teorema nos indica que $\neg(\forall\mu\in[\hbox{\O}]_n)(\mu=0)$, lo cual concluye la prueba.}
{\theorem Si $n\leq m\Rightarrow[\hbox{\O}]_n\subseteq[\hbox{\O}]_m$.}\\
{\proof Si $\mu\in[\hbox{\O}]_n\Rightarrow \mu^n=0$, luego $\mu^m=\mu^n*\mu^{m-n}=0$, entonces $\mu\in[\hbox{\O}]_m$.}
{\theorem Si $\mu\in [\hbox{\O}]_n \Rightarrow \forall x\neq 0$, $x*\mu\in[\hbox{\O}]_n$.}\\
{\proof $(x*\mu)^n=(x*\mu)*...(x*\mu)*(x*\mu)=x^n*\mu^n=0$.}
{\theorem $\MPm \vdash\ \kk/\sim_n$ es un campo algebraico.}\\
{\proof Dado cualquier valor en $\kk$ este puede tener la forma $x+\mu$, luego solo la parte $x$ (con $x^n\neq 0$) pertenece a $\kk/\sim_n$, siendo $x=0$ el \'unico valor en $\kk/\sim_n$ que cumple $x^n=0$. Por consiguiente, si $(x\neq 0)\wedge(x\in \kk/\sim)$, entonces $(\exists y\in \kk/\sim\ |\ x*y=1)$. Adicionalmente $(\forall x,y\in\kk/\sim) \Rightarrow x*y=y*x)$, teniendo en cuenta el teorema \ref{T1}. Aqu\'{\i} $\kk/\sim$ se puede interpretar como el objeto $K$ de valores de la medici\'on, ya que $K$ tiene la propiedad de ser un campo algebraico (ver aparte No V).}
{\theorem Si $\mu,\eta\in[\hbox{\O}]_n$, $p,q,r,s,k\in\nn$, $0<(p;q;r;s)<n$ y $0<k$, entonces\label{pg}}
{\case $(\mu^p+\eta^q)^{2k}=(\mu^p*\eta^q)^k+(\eta^q*\mu^p)^k$, $n/2\leq(p;q)<n$.\label{pg-1}}
{\case $(\mu^p+\eta^q)^{2k+1}=(\mu^p*\eta^q)^k*\mu^p+\eta^q*(\mu^p*\eta^q)^k$, $n/2\leq(p;q)<n$.\label{pg-2}}
{\case $(\mu^p*\eta^q+\eta^r)^{k}=(\mu^p*\eta^q)^k+\eta^r*(\mu^p*\eta^q)^{k-1}$, $q+r\geq n$.\label{pg-3}}
{\case $(\mu^p*\eta^q+\mu^s)^{k}=(\mu^p*\eta^q)^k+(\mu^p*\eta^q)^{k-1}*\mu^s$, $p+s\geq n$.\label{pg-4}}
{\case $(\mu^p*\eta^q+\eta^r*\mu^s)^{k}=(\mu^p*\eta^q)^k+(\eta^r*\mu^s)^{k}$, $(q+r;p+s)\geq n$.\label{pg-5}}\\
{\proof El c\'alculo algebraico es directo, o por inducci\'on para $k$.}
{\corollary Si $r=q$, $s=p$ en el teorema \ref{pg}, se tiene que\label{corog}}
$$(\mu^p+\eta^q)^{2k}=(\mu^p*\eta^q)^k+(\eta^q*\mu^p)^k=(\mu^p*\eta^q+\eta^q*\mu^p)^{k}$$
{\proof Directo de \ref{pg-1} y \ref{pg-5}.}
{\theorem Si $\mu,\nu\in[\hbox{\O}]_2$, se tiene que}
{\case $(\mu+\nu)^{2k}=(\mu*\nu)^k+(\nu*\mu)^k$.\label{ctpoli-1}}
{\case $(\mu+\nu)^{2k+1}=(\nu*\mu)^k*\nu+\mu*(\nu*\mu)^k$.\label{ctpoli-2}}
{\case $(\mu*\nu+\mu)^{k}=(\mu*\nu)^k+(\mu*\nu)^{k-1}*\mu=(\mu*\nu)^{k-1}*(\mu*\nu+\mu)$.\label{ctpoli-3}}
{\case $(\mu*\nu+\nu)^{k}=(\mu*\nu)^k+\nu*(\mu*\nu)^{k-1}=(\mu*\nu+\nu)*(\mu*\nu)^{k-1}$.\label{ctpoli-4}}
{\case $(\mu*\nu+\nu*\mu)^{k}=(\mu*\nu)^k+(\nu*\mu)^k$.\label{ctpoli-5}}\\
{\proof tome $p=q=r=s=1$ para cada uno de los casos en el teorema \ref{pg}.}
{\definition Se define el conmutador $[,]$ y el anticonmutador $\{,\}$ como} 
$$
[x,y]=x*y-y*x\qquad \{x,y\}=x*y+y*x.
$$
Es obvio que $2x*y=[x,y]+\{x,y\}$, pero a diferencia de los n\'umeros reales standard, 
{\theorem Si $\ \mu,\nu\in[\hbox{\O}]_n$, se tiene que}
{\case $\{\mu^p,\nu^q\}=0\quad\Leftrightarrow\quad\mu^p+\nu^q\in[\hbox{\O}]_{2}$.}\\
{\proof Tome $k=1$ en el corolario \ref{corog}.}
{\case Para $n$ par, $\mu^p+\nu^q\in[\hbox{\O}]_{n}\Leftrightarrow  \{\mu^p,\nu^q\}\in[\hbox{\O}]_{n/2}$.}\\
{\proof Directo del corolario \ref{corog} con $n=2k$.}
{\case $n$ impar, $\mu^p+\nu^q\in[\hbox{\O}]_{n}\Leftrightarrow(\mu^p\nu^q)^{(n-1)/2}*\mu^p+\nu^q*(\mu^p*\nu^q)^{(n-1)/2}=0$.}\\
{\proof Directo del caso \ref{pg-2} con $n=2k+1$.}
{\case $\mu^p*\eta^q,\eta^q*\mu^p\in[\hbox{\O}]_n\Rightarrow\{\mu^p,\eta^q\}\in [\hbox{\O}]_n$.}\\ 
{\proof Tome $k=n$ en el corolario \ref{corog}.}
{\theorem Si $\ \mu,\nu\in[\hbox{\O}]_2$, se tiene que}
{\case $\mu+\nu\in[\hbox{\O}]_2\quad\Leftrightarrow\quad\{\mu,\nu\}=0$.}
{\proof Tome $k=1$ en \ref{ctpoli-1}.}
{\case $\mu*\nu,\nu*\mu\in[\hbox{\O}]_2\ \Rightarrow\ \{\mu,\nu\}\in[\hbox{\O}]_2$.}
{\proof Tome $k=2$ en \ref{ctpoli-5}}
$$
\begin{array}{l}
\hbox{En resumen, si}\\
\mu,\nu,\mu+\nu\in[\hbox{\O}]_2\\
x,y\in{\kk}/\sim_2
\end{array}
\quad \Rightarrow \quad
\begin{array}{c}
[x,y]=0       \qquad \{x,y\}=2(x*y)\\
\{\mu,\nu\}=0 \qquad [\mu,\nu]=2(\mu*\nu)\\
\left[ x+\mu,y+\nu\right] = [\mu,\nu]
\end{array}
$$
En general si 
$$
\mu\in[\hbox{\O}]_n,\quad x\in{\kk}/\sim_n, \quad\Rightarrow\quad [x,\mu]=0
$$
{\theorem $\MPm\vdash(\forall f\in\kk^\kk)$, $f$ tiene una extensi\'on en $[\ ]_2$.\footnote{Usaremos $[\ ]_2$ para indicar una m\'onada de orden 2.}}\\
{\proof $\phi(\mu)=f(x+\mu)=f(x)+\mu*f'(x)$, luego si $f(x)+\mu*f'(x)=f(x+\mu)=f(x)+\mu*f''(x)\Rightarrow f'(x)=f''(x)$.}
Teniendo en cuenta que $f(x+\mu)-f(x)=\mu*f'(x)$, podemos definir {\sf la extensi\'on} ($\Df$) en $x$ para cualquier funci\'on $f$, luego
{\definition $\mu*\Df f(x)=f(x+\mu)-f(x)=\mu*f'(x) \Rightarrow \Df f(x)=f'(x).$}
{\theorem Si $\mu\in[\hbox{\O}]_{n+1}$, $\mu^n*f(x+\mu)=\mu^n*f(x)$.} {\proof $\mu^n*f(x+\mu)=\mu^n*f(x)+\mu^{n+1}*f'(x)+\mu^{n+2}*f''(x)+...=\mu^n*f(x)$.} 
{\theorem Para $\mu\in[\hbox{\O}]_{n+1}$, se tiene que} 
$$
\mu^{n-1}*(f(x+\mu)-f(x))=\mu^{n}*\Df f(x).
$$
{\proof Inmediato, ya que $\mu^{n-1}*f(x+\mu)=\mu^{n-1}*f(x)+\mu^{n}*f^{(1)}(x)$.}
{\definition Para $p\leq n$ y $\mu\in[\hbox{\O}]_{n+1}$, tomando $\Df^0f(x)=f(x)$. Se define de manera inductiva {\sf la p-extensi\'on} de $f$ en $x$ como
$$
\mu^n*\Df^pf(x)=\mu^{n-p}*\left(f(x+\mu)-\sum_{i=0}^{p-1}\mu^i \Df^{i}f(x)\right)\Rightarrow \Df^{p}f(x)=f^{(p)}(x).
$$}
Todo esto podr\'{\i}a ser interpretado como que: toda funci\'on de $\kk\to\kk$ tiene una expansi\'on en serie de Taylor. Pero se debe tener cuidado ya que las variables $\mu$ no necesariamente tiene interpretaci\'on de variaci\'on infinitesimal (es decir $x+\delta x=y$, con $\|x-y\|\neq 0$), dado que $[x]=[x+\mu]=[x]+[\mu]=x+[\mu]$. Es decir no hay un movimiento neto sobre la m\'onada.\footnote{Usaremos $[\ ]$ para indicar una m\'onada de orden $n\geq 2$.} Adem\'as las variables internas no necesariamente son variables conmutativas, como sucede en las m\'onadas de orden 2. Es interesante notar que para este caso las m\'onadas se comportan como variables de Grassmann, las cuales son ampliamente conocidas y estudiadas en la f\'{\i}sica te\'orica. Estas implican una nueva simetr\'{\i}a o supersimetr\'{\i}a en las ecuaciones de campo, lo cual es \'util cuando se quiere una TCC libre de infinitos y que permita una teor\'{\i}a cu\'antica de la gravedad. Siendo esta la \'unica ruta que se ha conseguido con este objetivo. 

Es aqu\'{\i} donde se revela uno de los prop\'ositos al formular una f\'{\i}sica mon\'adica, ya que nuestro inter\'es es encontrar una f\'{\i}sica sin infinitos y que posibilite una teor\'{\i}a cu\'antica de la gravedad, adem\'as de un contenido matem\'atico intuitivo. Luego, el primer criterio en esta direcci\'on es englobar lo que hasta ahora se conoce relativamente bien: las variables de Grassmann. Lo interesante de las m\'onadas nilpotentes es que estas estructuras posibilitan construir un c\'alculo diferencial con nuevas medidas, y en donde el lenguaje matem\'atico se vuelve m\'as rico en estructuras y en posibilidades.

\begin{flushright}
{\it Posibilidad e imposibilidad del experimento mental y de la f\'{\i}sica te\'orica.} \bf{(XII)}
\end{flushright}

El experimento mental (o ``Gedanken experiment'') ha tenido gran importancia en la f\'{\i}sica moderna desde Galileo hasta el presente con el famoso ``experimento mental de Einstein-Rosen-Podosky''. Su uso es indispensable en el momento de criticar, o de poner a prueba una teor\'{\i}a o un modelo. En un ``experimento mental'' se parte de un supuesto $P$ que es desarrollado para inferir $Q$ ($P\Rightarrow Q$), que o bien debe ser corroborado o indica una contradicci\'on. Si lo \'ultimo ocurre, se puede tener que $P$ no es cierto, o nuesto m\'etodo de argumentaci\'on tiene deficiencias, o ambos.

Sin embargo, el ``experimento mental'' debe ser modificado dentro de un esquema f\'{\i}sico basado en un principio de acci\'on constructiva, debido a que la f\'{\i}sica ahora tiene ligaduras que antes no teniamos y el proceso de idealizaci\'on debe ser m\'as cuidadoso: es necesario mostrar que existe la posibilidad f\'{\i}sica de obtener los objetos de la argumentaci\'on. Luego no se puede dise\~nar un proceso f\'{\i}sico $P$ si no es posible realizarlo o contrastarlo, lo cual acota la posibilidad de conjeturar. De otra forma, el ``experimento mental'' no escapa a las indicaciones del {\sf PAC}. Entonces, argumentos hipot\'eticos que nunca han sido o que no pueden ser contrastados no pueden ser considerados como f\'{\i}sicos. Aunque, esto no quiere decir que deje de tener sentido acad\'emico, como de echo ocurre.

Por consiguiente, en f\'{\i}sica mon\'adica nuestra imaginaci\'on f\'{\i}sica ahora tiene liga-duras que antes no teniamos. Por ej. no tiene sentido f\'{\i}sico considerar adjetivos como: perfecto, exacto, punto, instante, infinito, incertidumbre nula, etc, debido a que estos son idealizaciones que no pueden ser perceptibles, ni contrastables. Sin embargo, las consecuencias del {\sf PAC} son aun m\'as delicadas, porque se pierden muchos conceptos geom\'etricos debido a que estos son idealizaciones o juegos de la mente que dificilmente tienen un contraste f\'{\i}sico, y que solo sirven para ayudarnos a entender. Por ejemplo, esto $\bigtriangleup$ o esto $\bigcirc$, no es ni un tri\'angulo, ni una circunferencia,\footnote{recuerde: estos tienen un espesor, y gracias a esto usted puede ver estas figuras, adem\'as de que con un lupa lo suficientemente potente, el lector podr\'a comprobar que pierden las caracter\'{\i}sticas que los definen.} estos habitan en su mente, y son solo representaciones de lo que usted cree que son. 

Todo esto podr\'{\i}a llevar a la desesperaci\'on a quien necesita de figuras geom\'etricas en la f\'{\i}sica, pero es posible rescatar estos conceptos si se introduce el concepto de {\sf aproximado} o {\sf localmente cierto}. Es decir, se puede recuperar muchas estructuras geom\'etricas dependiendo de la escala con la que se miren, es decir con la escala que se midan. Por ejemplo a escala microsc\'opica deja de tener sentido geom\'etrico euclideano esto $\bigtriangleup$, pero si aumentamos la escala (la cual depende del grosor del trazo) obtenemos que estos pueden dar una imagen aproximada de lo que queremos decir. De otra forma, los objetos dependen del contexto en que son enunciados, o de como se les ``miren'', lo que nos lleva a introducir el concepto de {\sf mutabilidad}: ``un objeto es lo que ES, dependiendo de como se le observe''.

En f\'{\i}sica siempre se busca que todo problema complejo se pueda dividir en proble-mas simples que nos permiten abordar y solucionarlo. Sin embargo estas aproximaciones a veces pueden ser solo condiciones ideales que no tienen sentido f\'{\i}sico, por ej. puntos equilibrio inestable, funciones discontinuas o no derivables en alg\'un punto, etc. Y aunque estos no son f\'{\i}sicos, son herramientas provisionales a nuestro entendimiento. Pero, al reconstruir la f\'{\i}sica siguiendo las indicaciones del {\sf PAC}, tenemos cotas para el uso del lenguaje matem\'atico. Como es el caso, en un modelo nilpotente conmutativo de la f\'{\i}sica mon\'adica toda funci\'on tiene una extensi\'on suave, luego no se puede considerar curvas que son discontinuas o que por ejemplo: no sean derivables en alguno de sus elementos.

Por otro lado, pierde sentido las estructuras fractales en la f\'{\i}sica mon\'adica, debido a que estos objetos solo se obtienen de manera recursiva y como un l\'{\i}mite al infinito. Adem\'as de que posibilitan una estructura geom\'etrica no-trivial a cualquier escala o resoluci\'on, lo cual no tiene sentido en la f\'{\i}sica, porque siempre tendremos una incertidumbre y aun m\'as si esta es irresoluble. Sin embargo, objetos similares a estos se recuperar\'{\i}an en {\MP}, definiendo un {\sf fractal de orden n}: como ``el objeto de la n-esima iteracci\'on''. En esta misma direcci\'on, tampoco tiene mucho sentido considerar un sistema din\'amico ca\'otico tal como est\'a definido en la actualidad\footnote{Un sistema din\'amico $F$ es ca\'otico si 
\begin{itemize}
\item Los puntos peri\'odicos para $F$ son densos en el espacio de fase ${\mathcal G}$: Dado un punto cualquiera $x$ en el espacio de fase ${\mathcal G}$, existe un punto peri\'odico tan cercano a $x$ como se quiera.
\item $F$ depende sensitivamente de las condiciones iniciales: Dado un $\beta>0$, tal que para cualquier $x$ y cualquier $\varepsilon >0$, existe un $y$ cercano a $x$ con $d(x,y)<\varepsilon$ \& $k$, tal que $d(F_k(x),F_k(y))\geq\beta$.
\item $F$ es transitivo: Para cualquier dos puntos $x,y$ existe un $z$ con $d(x,z)<\varepsilon$ cuya \'orbita pasa por $y$ a una distancia menor de $\varepsilon$ (es decir $d(y,F_n(z))<\varepsilon$ para alg\'un $n$).
\end{itemize}}. Ya que por la misma raz\'on antes expuesta, no se puede comprobar en un laboratorio las propiedades que lo definen. Con esto no estoy diciendo que el caos no exista en la f\'{\i}sica y que deje de tener inter\'es acad\'emico; porque de echo lo tiene. Lo que se quiere plantear, es la diferencia entre lo que es matem\'aticamente coherente \& lo que es f\'{\i}sicamente perceptible y permisible, la diferencia entre experimento \& abstracci\'on. Y se debe reconocer que cualquier abstracci\'on nunca representar\'a al universo f\'{\i}sico dado que; por un lado no se puede tener una ``imagen'' acabada de la realidad, y por otro lado no se puede tener una correspondencia 1-1 entre la experiencia y su descripci\'on. Luego, cualquier sistema formal solo tendr\'a sentido para efectos pr\'acticos de representaci\'on y comunicaci\'on.

\begin{flushright}
{\it Agradecimientos.}
\end{flushright}

El autor fue parcialmente auspiciado como becario por {\it la Fundaci\'on Mazda para el Arte y la Ciencia}, instituci\'on a la cual \'el expresa su gratitud, e igualmente a Alvaro Obyrne por sus comentarios y sugerencias hechas a este documento.

\begin{flushright} 
{\it Este trabajo est\'a dedicado a aquellos seres cercanos y m\'as queridos\\
que me han acompa\~nado dia a dia en el teatro de la vida.}
\end{flushright}

El siguiente contenido bibliogr\'afico es un complemento y no una referencia a este documento. Donde el lector puede consultar algunas apartes mencionados, pero en m\'as detalle y profundidad. La etiqueta entre corchetes $[\ ]$, indica la secci\'on asociada.

\end{document}